\documentclass[aip,apl,reprint,amsmath,amssymb]{revtex4-1}
\usepackage[english]{babel}
\usepackage{graphicx}
\draft 

\usepackage[usenames,dvipsnames]{xcolor}
\usepackage{color, colortbl}
\usepackage{float} 
\definecolor{Gray}{gray}{0.9}
\definecolor{LightC}{rgb}{0.88,1,1}
\definecolor{LightY}{rgb}{1,0.9,0.7}
\definecolor{LightR}{rgb}{1,0.7,0.7}
\definecolor{LightRo}{rgb}{1,0.7,0.9}
\definecolor{LightV}{rgb}{0.8,0.7,1}
\definecolor{LightG}{rgb}{0.7,1,0.7}
\definecolor{LightB}{rgb}{0.7,0.8,1}
\usepackage{xcolor}
\usepackage{soul}
\newcommand{\mathcolorbox}[2]{\colorbox{#1}{$\displaystyle #2$}}
\newcommand{\dd}{\mathrm{d}}

\begin{document}


\title{Nonlinear elasticity of silica nanofiber}


\author{Adrien Godet}
\affiliation{Institut FEMTO-ST UMR CNRS 6174, Universit\'e Bourgogne Franche-Comt\'e, Besan\c{c}on, France}

\author{Thibaut Sylvestre}
\affiliation{Institut FEMTO-ST UMR CNRS 6174, Universit\'e Bourgogne Franche-Comt\'e, Besan\c{c}on, France}

\author{Vincent P\^echeur}
\affiliation{Institut FEMTO-ST UMR CNRS 6174, Universit\'e Bourgogne Franche-Comt\'e, Besan\c{c}on, France}

\author{Jacques Chr\'etien}
\affiliation{Institut FEMTO-ST UMR CNRS 6174, Universit\'e Bourgogne Franche-Comt\'e, Besan\c{c}on, France}

\author{Jean-Charles Beugnot}
\affiliation{Institut FEMTO-ST UMR CNRS 6174, Universit\'e Bourgogne Franche-Comt\'e, Besan\c{c}on, France}

\author{Kien Phan~Huy}
\email[Email:]{kphanhuy@univ-fcomte.fr}
\affiliation{Institut FEMTO-ST UMR CNRS 6174, Universit\'e Bourgogne Franche-Comt\'e, Besan\c{c}on, France}


\date{\today}

\begin{abstract}
Optical nanofibers (ONFs) are excellent nanophotonic platforms for various applications such as optical sensing, quantum and nonlinear optics, due to both the tight optical confinement and their wide evanescent field in the sub-wavelength limit. Other remarkable features of these ultrathin fibers are their surface acoustic properties and their high tensile strength. Here we investigate Brillouin light scattering in silica-glass tapered optical fibers under high tensile strain and show that the fundamental properties of elastic waves dramatically change due to elastic anisotropy and nonlinear elasticity for strain larger than 2\%. This yields to unexpected and remarkable Brillouin strain coefficients for all Brillouin resonances including surface and hybrid waves, followed by a nonlinear evolution at high tensile strength. We further provide a complete theoretical analysis based on third-order nonlinear elasticity of silica that remarkably agrees with our experimental data. These new regimes open the way to the development of compact tensile strain optical sensors based on nanofibers.
\end{abstract}


\maketitle 


\section{Introduction} 
Optical micro and nanofibers (ONFs) are long and uniform ultra-thin fibers manufactured by heating and tapering standard optical fibers down to the sub-micrometer scale  \cite{birks_shape_1992,tong_subwavelength-diameter_2003,tong_book,brambilla_optical_2010,foster_nonlinear_2008}. In addition to providing a tight optical confinement, they exhibit a strong evanescent field in the sub-wavelength limit, which is very attractive for applications such as optical sensing and quantum photonics \cite{tong_book,brambilla_optical_2010,sayrin_storage_2015,gouraud_demonstration_2015}. They also possess exceptional mechanical and elastic properties, with large extensibility and high tensile strength \cite{Holleis_2014}. From an acoustic viewpoint, it has been recently shown that ONFs support new class of acoustic waves compared to standard fibers owing to the strong coupling between shear and longitudinal waves \cite{beugnot_brillouin_2014,florez_brillouin_2016}. These include hybrid acoustic waves (HAWs) and surface acoustic waves (SAWs) that move at a lower speed than the longitudinal acoustic velocity. This in turn gives rise to a multi-peaked Brillouin spectrum with Brillouin frequency shift ranging from 5 GHz for SAWs up to 10 GHz for HAWs.

In this work we investigate the tensile strain dependence of optical nanofibers using Brillouin spectroscopy and we show that the fundamental elastic properties dramatically change due to strain-induced elastic anisotropy. More specifically, we find that the Brillouin strain coefficients associated with SAWs and HAWs are completely different from that observed in standard optical fibers. We report the observation of unexpected Brillouin frequency shifts featuring different slopes and crossings. Unlike in standard fibers, they do not scale with the Brillouin frequency shift anymore and even show a strong nonlinear deviation at high tensile strain. To get further understanding, we develop a theoretical model based on third-order elasticity of silica to predict the strain dependence of all acoustic waves in tapered optical fibers including the nanofiber section and the transitions. We find a very good agreement with experimental results.

The article is divided into four sections. We briefly recall the story of Brillouin scattering and of the strain's law in the first section. The second section is devoted to the experimental measurements of Brillouin strain coefficients in optical nanofibers with different diameters and tensile train conditions. In a third section, we provide the analytic model based on the nonlinear Hooke's law to clearly interpret our observations. Finally, our results are further compared to finite element method (FEM)-based numerical simulations to highlight both the nanofiber and the transition contributions to the Brillouin spectrum. 

\section{History}
In 1925, two years after discovering the scattering of light through elastic waves in homogenous transparent materials \cite{brillouin_diffusion_1922}, L\'eon Brillouin theoretically predicted that the speed of elastic waves depends on strain due to the nonlinear elasticity  and later provided the magnitude and sign of the third-order elastic constants \cite{brillouin_les_1925,brillouin_les_1946}. These predictions were experimentally checked a few years later with ultrasonic velocity measurements \cite{hughes_second-order_1953}, and the first measurement of all third-order elastic constants of fused silica was reported in 1965 \cite{bogardus_thirdorder_1965}, followed by other measurements still using ultrasonic techniques \cite{Wang_temperature_1992}. Brillouin light scattering was later exploited for measuring the longitudinal and transverse velocities in many bulk materials \cite{pine_brillouin_1969,vacher_temperature_1975,pelous_thermal_1976}. In optical fibers, tensile strain measurements lead to the third-order elastic constant $C_{111}$\cite{horiguchi_tensile_1989}. All these fundamental research works have opened up the development of long range and high-resolution strain and temperature Brillouin distributed optical fiber sensors, that are today widely and commercially used for integrity and security in civil engineering and petroleum industry \cite{Galindez_2012}. The Brillouin-based fiber optical sensors principally exploit the sensitivity of the Brillouin gain spectrum (BGS) in single-mode fibers to temperature and strain \cite{Culverhouse_1989,horiguchi_1990,Nikles_1996}. However, given the dimension of a standard single-mode optical fiber (SMF) and the acoustic frequency involved in the Brillouin scattering, only the bulk longitudinal acoustic wave is probed by Brillouin sensors. The first consequence is that the dependence of the Brillouin frequency shift $\nu_B$ with respect to the tensile strain is 
\begin{equation}
C_\epsilon=\frac{\dd \nu_B}{\dd \epsilon}=\frac{1}{2}\frac{\nu_B}{C_{11}}C_{111}
\label{eq:01}
\end{equation}
 where $C_{11}$ and $C_{111}$ are the longitudinal second-order and third-order elastic constants, respectively (See Table \ref{tab:1}), $\epsilon$ is the tensile strain, usually defined in percentage of fiber elongation and the refractive index dependence to $\epsilon$ is neglected. For instance, in standard optical silica fibers, the Brillouin frequency shift is about 11 GHz and thus the strain coefficient is $C_\epsilon \simeq 0.04~MHz/\mu\epsilon$ at a wavelength of 1550 nm which can also be written 400 MHz/\% \cite{Galindez_2012}.  Eq. (\ref{eq:01}) generally assumes that only longitudinal mechanical properties are sensed by the apparatus. In standard optical fibers, the maximal tensile strain is limited to 2\% due to the polymer cladding but naturally the elasticity of silica glass allow going up to 6\%.\cite{gueretteNonlinearElasticitySilica2016} 

In this work, we show that this fundamental assumption is no longer valid in optical nanofibers that carry both surface acoustic waves and hybrid acoustic waves \cite{beugnot_brillouin_2014}. More specifically, we show that, when applying a tensile strain on a tapered optical fiber, the nonlinear elasticity of the nanofiber induces a strong elastic anisotropy that in turn affects the acoustic wave propagation and the Brillouin strain coefficients. As we will see thereafter, these dramatic changes in the mechanical properties of the nanofibers significantly impact the Brillouin strain coefficients that do no scale anymore with the Brillouin frequency shift $\nu_B$.

\section{Experiments}
\begin{figure} 
\includegraphics[width=8cm]{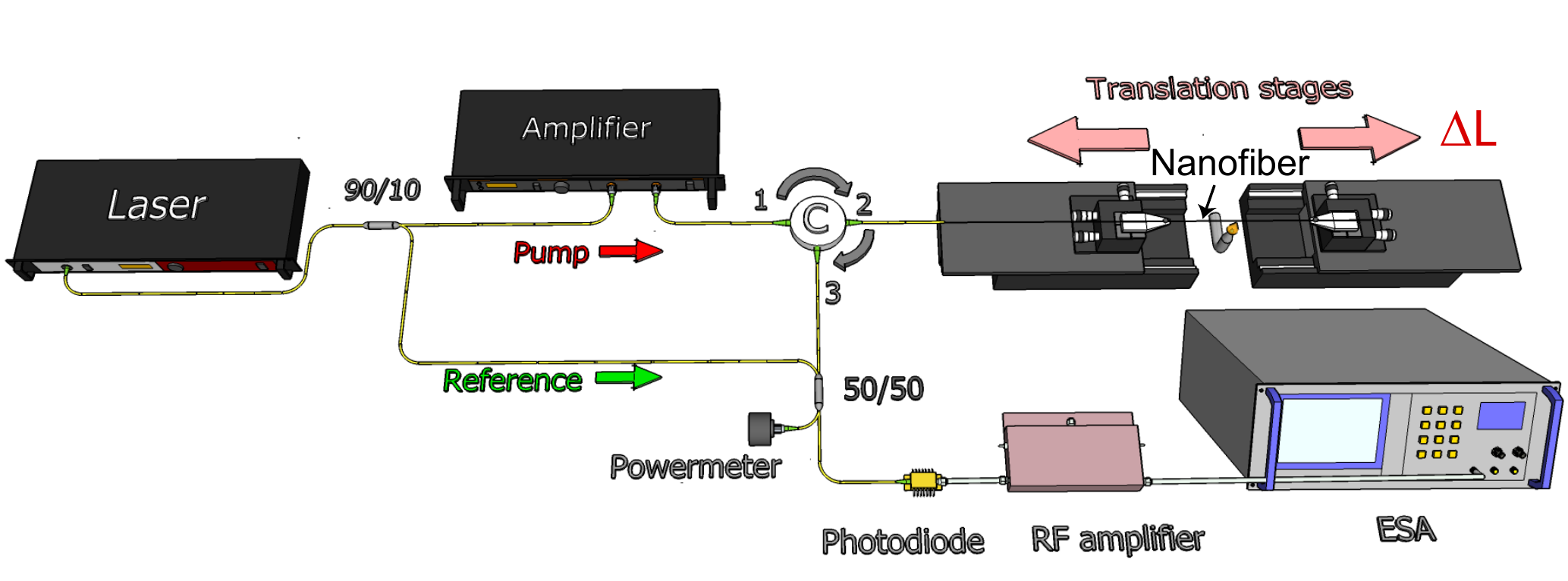}
\caption{Experimental setup for measuring the backward Brillouin spectrum of tapered optical fibers using heterodyne detection. Laser: Coherent distributed-feedback (DFB) laser at 1550 nm, Amplifier: 33 dBm EDFA, C: optical circulator, ESA: electrical spectrum analyser. As the translation stages move apart, the tapered optical fiber is stretched by $\Delta L$.}\label{fig:Setup}            
\end{figure}

Figure \ref{fig:Setup} shows a scheme of the experimental setup for making, stretching and characterizing the optical nanofibers using Brillouin spectroscopy. The ONFs were fabricated by heating and tapering un-coated standard single-mode fibers (SMFs) using the heat-brush technique \cite{godet_brillouin_2017}. Summing up, this technique uses a stabilized flame and two motion-controlled translation stages to heat and stretch the fibers. The nanofiber length and taper transition shapes were fully controlled by the computed trajectories of the two translation stages, while keeping the butane flame motionless. Using this technique, we were able to achieve ultra-thin fiber waists down to 600 nm over uniform length up to 80 mm. The transitions were adiabatically drawn to ensure single-mode conversion from the SMF to the nanofiber with nsertion loss down to 1.2 dB at 1550 nm. To probe the Brillouin backscattering signal, we used the heterodyne coherent detection technique shown in Fig \ref{fig:Setup}. This is a quite simple all-fiber setup where the light coming from a coherent continuous-wave laser at a wavelength of 1550 nm is split into two beams using a 90/10 fiber coupler. One beam is sent to an erbium-doped fiber amplifier (EDFA) and used as pump wave and the other beam is used as reference light. The pump light is injected through an optical circulator into the nanofiber under test and the backscattered Brillouin signal from the nanofiber is then mixed with the reference light using another 50/50 coupler. The two frequency-detuned beams give rise to an optical beat note that is further detected in the radio-frequency (RF) domain by a fast photodiode. The RF signal is then amplified and the resulting Brillouin spectrum is recorded with an electrical spectrum analyzer (ESA), once the pulling process is ended and the flame heat is dissipated. This prevents from any pollution, stress, and temperature effect that could shift the Brillouin resonances and alter the measurement results. To further investigate their nonlinear elastic properties, the Brillouin spectrum was measured by applying different tensile strains, still using the motion-controlled translation stages.  An exceptional axial extensibility up to 6\% was achieved before failure, while standard SMFs are usually limited to 2\% \cite{Galindez_2012}. 

\begin{figure} 
\includegraphics[width=8cm]{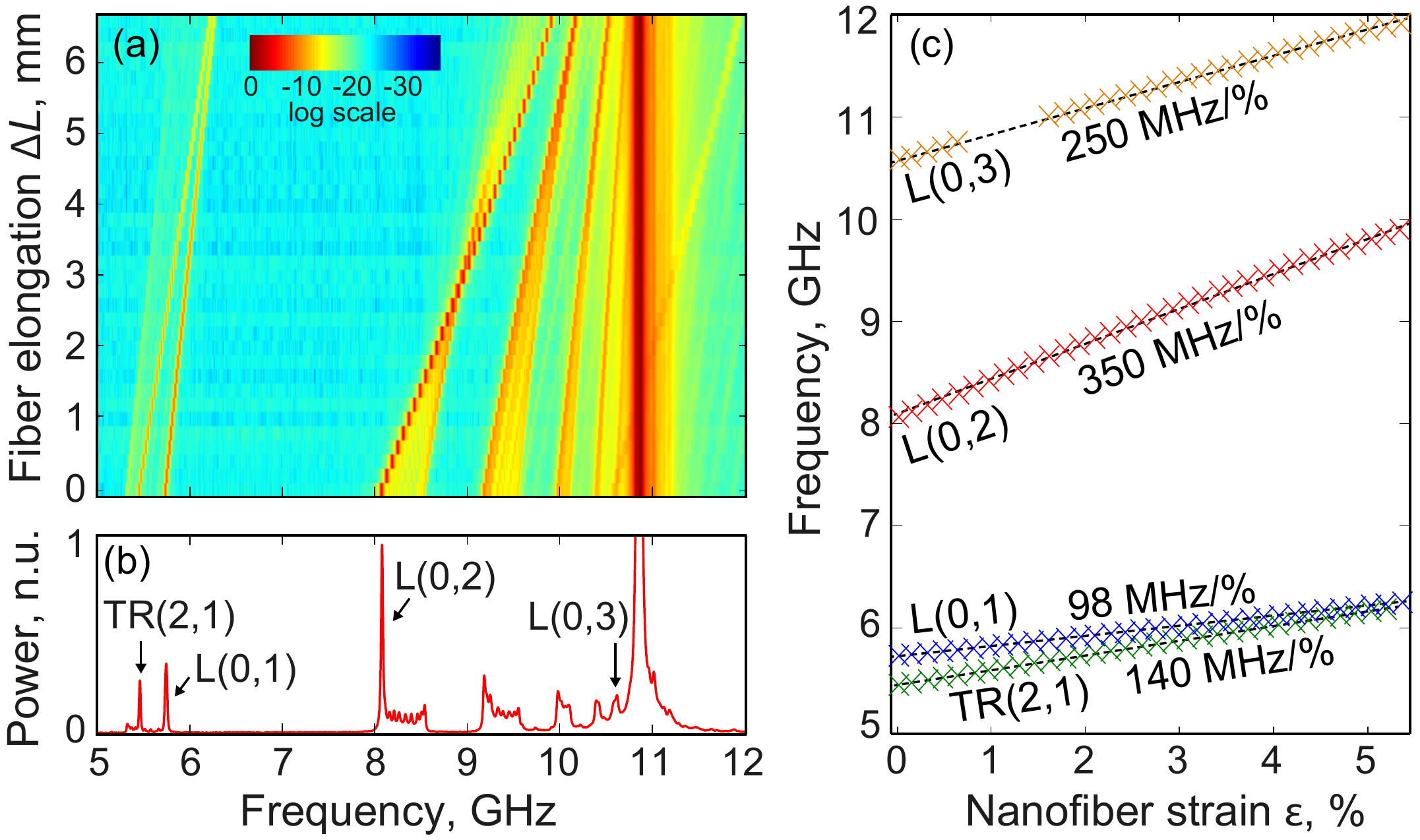}
\caption{(a) Colormap of the experimental Brillouin spectra measured in an optical nanofiber of diameter $d_{MF,1}$=660 nm and a length of 80 mm as a function of fiber tensile strain $\Delta L$ (vertical axis). (b) Experimental Brillouin spectrum (red curve) of the optical nanofiber without axial strain. The main acoustic modes in the nanofiber section are indicated by black arrows.  (c) Brillouin frequency shifts for each acoustic mode as a function of axial strain in percentage.}\label{fig:exp660}            
\end{figure}

Figure \ref{fig:exp660}(b) shows a typical experimental Brillouin spectrum in red, measured for an unstrained ONF with a diameter of 660-nm and a length of 80 mm. Note that these parameters were precisely estimated using the recent technique described by Beugnot \textit{et al.} \cite{godet_brillouin_2017}. As can be seen, the spectrum exhibits many frequency peaks in the frequency range 5-12 GHz, that are actually the signature of all the Brillouin acoustic resonances from both the tapered and un-tapered fiber sections. As previously shown in Ref. \cite{beugnot_brillouin_2014}, these include the surface (SAWs) and hybrid acoustic waves (HAWs) in the uniform waist section, as well as the strong Brillouin resonance around 11 GHz due to long un-tapered fiber sections. To get better insight, we have labelled in Fig. \ref{fig:exp660} the main acoustic modes from the ONF section only. They are denoted by TR(2,1), L(0,1), L(0,2) and L(0,3), respectively, where TR and L stand for torso-radial and longitudinal acoustic modes. We note that, at this micrometer scale, the acoustic waves propagate as modes and therefore they result from coherent superposition of both the longitudinal and the shear waves. Furthermore, these acoustic modes possess their own dispersion relation giving rise to such multi-peaked Brillouin spectra, compared to standard single-mode fibers that usually exhibit a single peak around 11 GHz, as that observed just above the L(0,3) \cite{rowell_brillouin_1978,volkov_second-_2015}. It is also important to emphasize that the surface and hybrid acoustic waves travel at acoustic velocities lower than the pure longitudinal acoustic wave, which gives rise to lower Brillouin frequency shifts down to 5.5 GHz, which is imposed by the Rayleigh limit \cite{beugnot_brillouin_2014}.

\begin{figure}[ht]
	\centering
	\includegraphics[width=8cm]{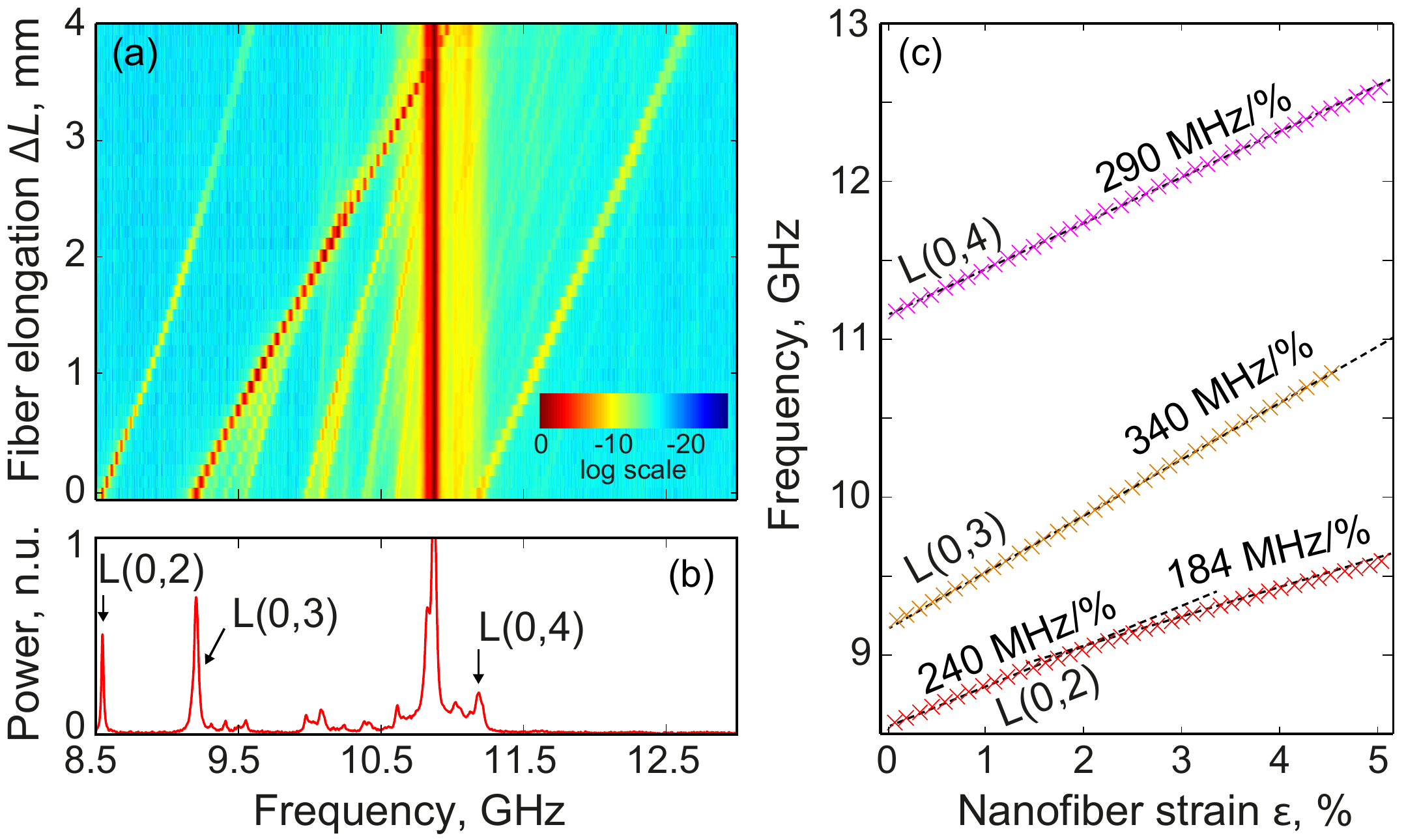}
	\caption{(a) Colormap of the experimental Brillouin spectra measured in an optical nanofiber of diameter $d_{MF,2}$=930 nm and a length of 40 mm as a function of fiber tensile strain $\Delta L$ (vertical axis). (b) Experimental Brillouin spectrum (red curve) of the optical nanofiber without axial strain. The main acoustic modes in the nanofiber section are indicated by black arrows.  (c) Brillouin frequency shifts for each acoustic mode as a function of axial strain in percentage.}
	\label{fig:OtherDiam}
	\end{figure}

Once the ONF is fabricated the flame is blown out. They are then characterized at room temperature and stretched by moving apart the translation stages by a distance $\Delta L$. This therefore strongly affects all the Brillouin acoustic resonances shifting them towards higher frequencies. This is shown in Fig. \ref{fig:exp660}(a) as a function of $\Delta L$ as a colormap. The Brillouin strain coefficients for all acoustic modes were further extracted from the slopes using the method detailed in Supplementary Informations \ref{A1} for the nanofiber strain computation. The data are reported in Fig. \ref{fig:exp660}(c) on which we can notice several striking features. Perhaps the most striking one is that the acoustic resonances do not scale with the Brillouin frequency $\nu_B$, as usually expected from Eq. (\ref{eq:01}). For instance, the acoustic mode L(0,2) has a steeper slope (350 MHz/\%, red crosses) than the L(0,3) modes (250 MHz/\%,orange crosses). Yet, what is most surprising is that the L(0,1) and TR(2,1) resonances cross each other for strain larger than $5\%$, just before failure. We would rather expected increasing slopes of the lines from bottom to top in Fig. \ref{fig:exp660}(c) but this is not the case. This proves that Eq. (\ref{eq:01}) is no longer valid for tapered fibers. To confirm these observations, we performed further experiments using other ONFs featuring different waist and length. Figures \ref{fig:OtherDiam}(a) and (b) show experimental results for a 930 nm diameter and 40-mm long nanofiber. Once again, the L(0,3) mode has a greater slope than the L(0,4), and the L(0,2) mode now features a strong nonlinear deviation versus the axial strain as shown in Fig. \ref{fig:OtherDiam}(c). 

	\begin{figure}[ht]
	\centering
	\includegraphics[width=8cm]{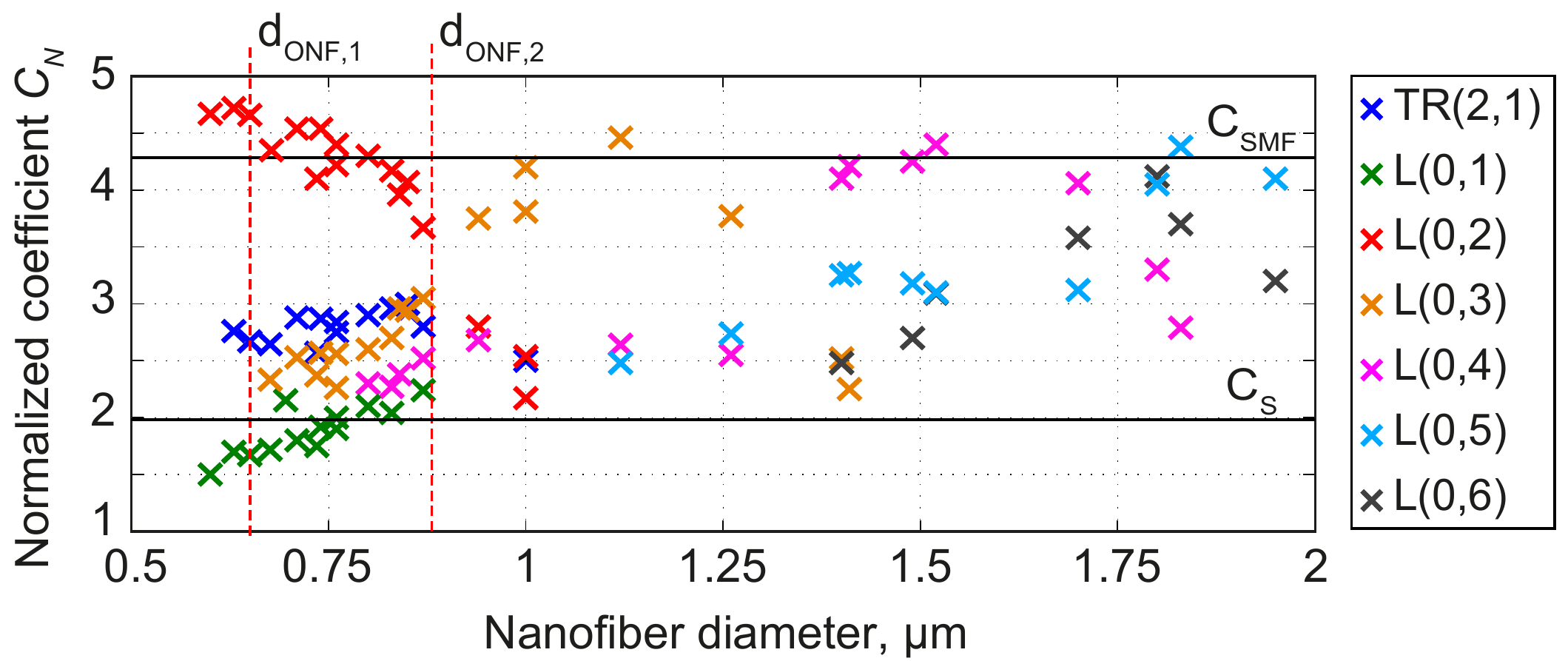}
	\caption{Normalized Brillouin strain coefficients $C_N$ of acoustic resonances for many fiber taper diameters ranging from 0.6 up to 1.9 $\mu$m (tensile strain $<$2\%). $C_{SMF}$ is the coefficient for SMF-28 related to longitudinal waves. $C_S$ refers to the computed coefficient using the acoustic shear velocity.}
	\label{fig:NormC}
	\end{figure}

We then repeated the same experiment using many taper waists ranging from 0.6 up to 1.9 $\mu$m. Figure \ref{fig:NormC} shows the normalized Brillouin strain coefficients measured for various acoustic resonances, defined as $C_N=\frac{1}{\nu_B}\frac{\dd \nu_B}{\dd \epsilon}$. For comparison, that of an SMF-28 is plotted as a horizontal black line (normalized coefficient $C_{SMF}=4.25$). According to Eq. (\ref{eq:01}), it corresponds to a pure longitudinal acoustic wave while the other bottom black line corresponds to the strain coefficient for a pure shear wave \cite{Tanaka_1999}. As can be seen, the coefficients of most nanofibers (color crosses) are highly dispersed in between these two limits and even beyond for small taper diameters less than 750 nm. These experimental data quite obviously show the limit of the standard model. In the next section, we will present a modified theory of tensile strain based on third-order nonlinear elastic constants that allows for predicting the Brillouin strain coefficients observed in nanofibers.
	
\section{Theory}
\begin{figure}[ht]
	\centering
	\includegraphics[width=8cm]{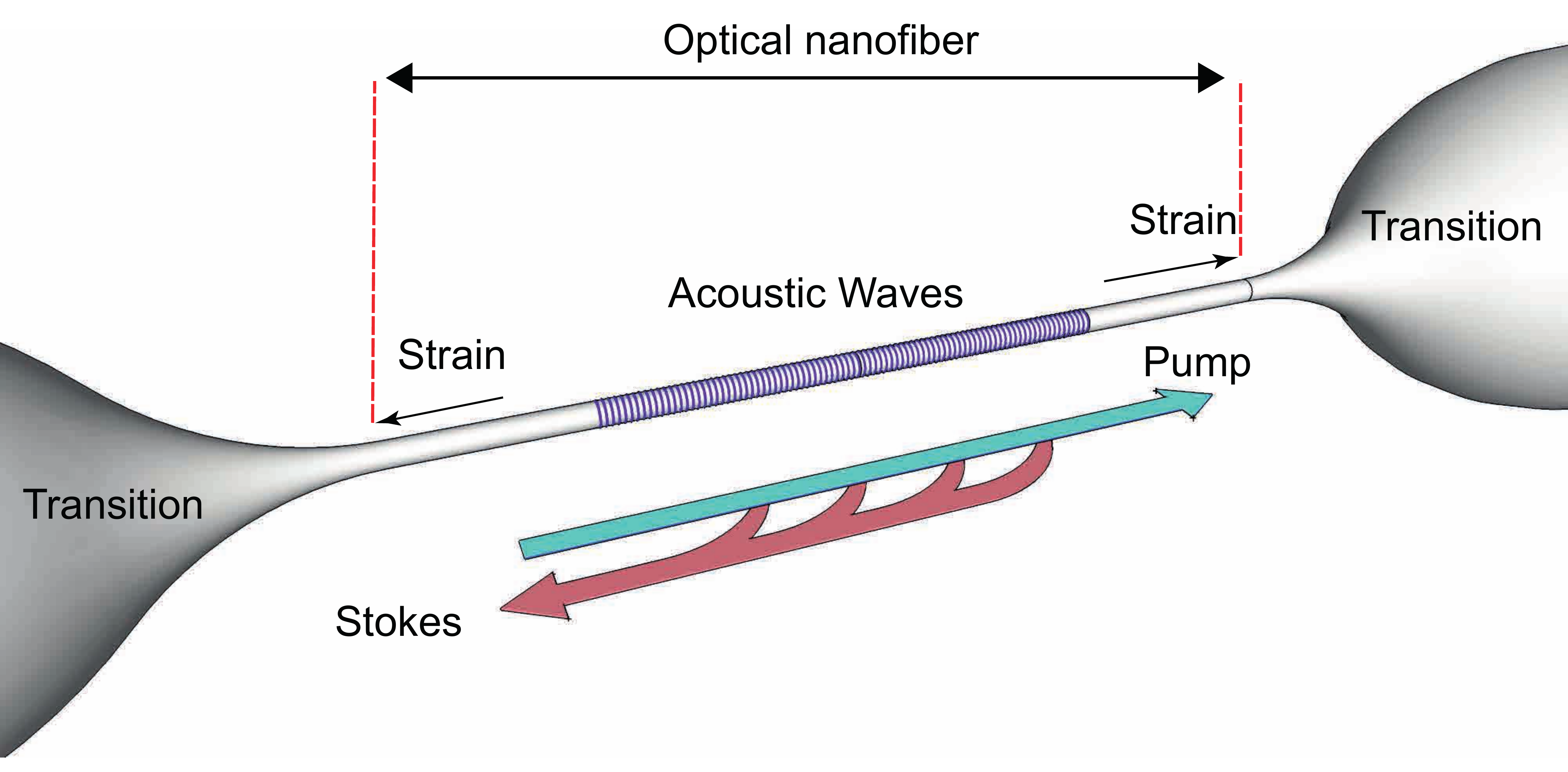}
\caption{The principle scheme of backward Brillouin scattering in an optical nanofiber. Blue arrow: Pump beam. Red arrow: Stokes beam. Purple: acoustic wave. When applying tensile strain the acoustic wave behavior dramatically changes due to elastic anisotropy. }
\label{fig:Principe}
\end{figure}There are four main reasons that can explain the unexpected Brillouin strain coefficients observed in Figs. \ref{fig:exp660}, \ref{fig:OtherDiam} and \ref{fig:NormC}. The first reason comes from the large extensibility of nanofibers. Tensile strain going up to 6\% leads to dramatic changes of their dimensions that can significantly affects the Brillouin resonances. Next the hybrid nature of the acoustic modes generated in ONFs gives rise to a strong nonlinear dispersion of the Brillouin resonances. The third reason relies on the nonlinear elasticity of silica which requires to consider all the third-order elastic tensor coefficients (TOEC). This is fundamentally different from Eq. (\ref{eq:01}) in which only the longitudinal third-order elastic coefficient $C_{111}$ was assumed. At last, when tensile strain is applied onto a material it acquires an elastic anisotropy that strongly affects the acoustic wave propagation. To elucidate this question, let us first recall the basics of backward Brillouin scattering (BS) in optical fibers, which arises from the interaction between two counter-propagating and frequency-detuned optical waves, and a longitudinal elastic wave, as shown schematically in Fig. \ref{fig:Principe} \cite{kobyakov_stimulated_2010}. This inelastic scattering is usually modelled using three coupled-amplitude equations for the pump wave (blue), the backward Stokes wave (red) and the elastic wave (purple)\cite{boydNonlinearOpticsThird2008}. Assuming the latter as a plane wave, this 3-wave interaction satisfies a phase-matching condition that reads as $\beta_a=\beta_p-\beta_S$. Neglecting the very small dispersion between the pump and Stokes waves, as $|\beta_p|\simeq|\beta_S|$, the phase-matching relation can be approximated by $|\beta_a|\simeq|2\beta_p|$, leading to the Brillouin frequency shift $\nu_B\simeq v\beta_p/\pi$, where $v$ is the velocity of the acoustic wave. This is worth emphasizing here that the acoustic velocity $v$ represents the main contribution to the strain effect \cite{horiguchi_tensile_1989}.

Unlike single-mode optical fibers, optical nanofibers have air cladding and a very small core at the scale of both the optical and acoustical wavelengths. As such, the elastic waves involved in Brillouin scattering can no longer be considered as plane waves, but as acoustic modes that include both longitudinal and shear elastic components. A direct consequence is that the acoustic velocity $v$ must be computed for each acoustic mode. To do so, we assume the displacement field as the superposition of a longitudinal wave of velocity $v_L$ and a shear wave of velocity $v_s$. Then we calculate the stress field using Hooke's law. In air, all stress components applied to the boundary of the ONF must be zero. Those boundary conditions eventually lead to a dispersion equation whose solutions are the acoustic velocities. However, the computation of the acoustic velocities is here a bit complex, since the tensile strain can reach up to a few percent, which is far beyond the linear elasticity regime. 

If the medium is strained so that infinitesimal deformations are no longer considered, nonlinear acoustic effects occur and the third-order elastic constants must be considered. The nonlinear Hooke's law can be written as \cite{volkov_second-_2015}
\begin{equation}
\sigma_{\alpha}=C_{\alpha\beta}\epsilon_{\beta}+\frac{1}{2}C_{\alpha\beta\gamma}\epsilon_{\beta}\epsilon_{\gamma},\label{eq:02}
\end{equation}
where $\sigma_{\alpha}$ is the $\alpha$ component of the stress tensor, $\epsilon_{\beta}$ is the $\beta$ component of the strain tensor, $C_{\alpha\beta}$ is the $(\alpha,\beta)$ component second-order elastic constant and $C_{\alpha\beta\gamma}$ the $(\alpha,\beta,\gamma)$ third-order elastic constant, with $\alpha,~\beta,~\gamma \in (rr, \theta \theta, zz,rz,\theta z, r\theta) \leftrightarrow (1,2,3,4,5,6)$ in cylindrical coordinates. 
The nonlinear regime has to be modelled with much attention, as we now have to solve jointly the static and dynamic solutions. The static solution is the amount of strain $\bar{\epsilon}_{zz}$ induced by the tensile strain, while the dynamic solutions are the acoustic waves $\tilde{\epsilon}_{\beta}$ that propagate along the strained optical microfibers. However, because of acoustic nonlinearity, the dynamic solutions are affected by the static solution at high tensile strain. In the following we will first solve the static problem by computing $\bar{\epsilon}_{zz}$ and we will further address the dynamic problem by computing the acoustic velocities for a given solution $\bar{\epsilon}_{zz}$. 

For the static problem, we can assume that only the longitudinal stress is significant and therefore we can write the nonlinear Hooke's law as $\bar{\sigma}_{zz}=c_{11}\bar{\epsilon}_{zz}+\frac{1}{2}c_{111}\bar{\epsilon}_{zz}^2$. Integrating the equation along the tapered fiber yields the expression of strain $\bar{\epsilon}_{zz}$ as a function of the total elongation \cite{Holleis_2014,Suhir_predicted_1993} (See SI \ref{A1}, for more details).

Before adressing the nonlinear dynamic problem, let us recall that ordinary linear elasticity and Hooke's law describe the physic of springs. It states that stress is proportional to deformation via the second-order elastic constants. One of the direct consequences of elasticity is the ability of a vibration to propagate as a wave. The wave velocity is then related to the second-order elastic constants. In a isotropic medium, there are twelve components in the elastic tensor and they result from a linear combination of two linearly independent coefficients, namely the Young and shear moduli, or the two Lam\'e constants ($\lambda,\mu$). As a result, in the linear regime, the second term of the right-hand side of Eq. \ref{eq:02} is negligible and the only term left is related to the stiffness matrix $C_{\alpha \beta}$. It is a symmetric tensor for an isotropic material, that can be written in Voigt notation as
\begin{equation}
	C_{\alpha \beta}\rightarrow C_0 = \left(
\begin{array}{cccccc}
 \cellcolor{LightY}\lambda+2\mu & \cellcolor{LightR} \lambda & \cellcolor{LightR} \lambda & 0 & 0 & 0 \\
 \cellcolor{LightR} \lambda & \cellcolor{LightY} \lambda+2\mu & \cellcolor{LightR} \lambda & 0 & 0 & 0 \\
 \cellcolor{LightR} \lambda & \cellcolor{LightR} \lambda & \cellcolor{LightY} \lambda+2\mu & 0 & 0 & 0 \\
 0 & 0 & 0 & \cellcolor{LightV}\mu & 0 & 0 \\
 0 & 0 & 0 & 0 & \cellcolor{LightV} \mu & 0 \\
 0 & 0 & 0 & 0 & 0 & \cellcolor{LightV} \mu \\
\end{array}
\right),
\end{equation}
where $\lambda=16$ GPa and $\mu=31$ GPa are the Lam\'e constants of fused silica. The colors in the matrix indicate the symmetry.
Solving the linear dynamic problem leads to two type of waves: the shear or longitudinal elastic waves.

In the nonlinear case, the tensile strain $\bar{\epsilon}_{zz}$ is strong enough to induce an acoustic nonlinearity. In the second term of the right-hand side of Eq. \ref{eq:02}, all terms involving $\bar{\epsilon}_{zz}$ are no longer negligible. The Hooke's law for the dynamic stress then reduces to 
\begin{equation}
\tilde{\sigma}_{\alpha}=C'_{\alpha\beta} \tilde{\epsilon}_{\beta}, \label{eq:02b}
\end{equation}
where 
\begin{equation}
C'_{\alpha\beta}=C_{\alpha\beta}+C_{\alpha\beta 3}\bar{\epsilon}_{zz}. \label{eq:03} 
\end{equation}
Next, to solve the dynamic problem we need to consider the strain-induced elastic anisotropy. In an isotropic medium, the matrix coefficients are a linear combination of three independent coefficients $C_{111}$, $C_{112}$, and $C_{123}$, respectively. As previously discussed, only the tensile strain $\bar{\epsilon}_{zz}$ is strong enough to provide an effective acoustic nonlinearity that breaks the isotropic symmetry of the tensor.  Therefore,  Eq. \ref{eq:03} only possesses $C_{\alpha \beta 3}$ third-order coefficients and rewrites the effective elastic tensor as follows

\begin{eqnarray}
&&C'_{\alpha \beta}=C_0\circ \nonumber\\ 
&&
\left\{
1
+
\left(
\begin{array}{cccccc}
 \cellcolor{LightY}3.05 & \cellcolor{LightR} 3.4 & \cellcolor{LightG} 15.05 & 0 & 0 & 0 \\
 \cellcolor{LightR} 3.4 & \cellcolor{LightY}3.05 & \cellcolor{LightG} 15.05 & 0 & 0 & 0 \\
 \cellcolor{LightG} 15.05 & \cellcolor{LightG}15.05 & \cellcolor{LightB} \textbf{6.71} & 0 & 0 & 0 \\
 0 & 0 & 0 & \cellcolor{LightRo}2.29 & 0 & 0 \\
 0 & 0 & 0 & 0 & \cellcolor{LightRo} 2.29 & 0 \\
 0 & 0 & 0 & 0 & 0 & \cellcolor{LightV}2.95 \\
\end{array}
\right)\bar{\epsilon}_{zz}
\right\},\label{eq:T}
\end{eqnarray}
where the data have been calculated from table \ref{tab:1} and $\circ$ denotes the Hadamard product. The tensor matrix is now typical of a medium with transverse isotropic symmetry with respect to the $z$ axis. As can be seen, the coefficients highlighted in blue and green give rise to a strong asymmetry in the other directions. For example, when the ONF is stretched at the maximum strain $\bar{\epsilon}_{zz}\simeq 6\%$, the elastic coefficient $C'_{12}$ significantly increases by more than $90\%$.

\begin{table}
\begin{tabular}{|l|l||l|l|}
\hline
   Refractive index $n$ & 1.444 & $C_{11}$ & 78 GPa\\
	Density $\rho$ & 2203 $kg/m^3$ &   $C_{44}$ & 31 GPa \\	
	$P_{11}$ & 0.12 & $C_{111}$ & 578 GPa \\
	$P_{12}$ & 0.27 & $C_{112}$ & 215 GPa \\ 
	$P_{44}$ & -0.073  & $C_{123}$ & 43 GPa\\
\hline
\end{tabular}
\caption{Optical and elastic parameters of fused silica used for nonlinear elasticity model and numerical simulations. $P_{ij}$ are electrostrictive tensor components. $C_{\alpha\beta}$ and $C_{\alpha\beta\gamma}$ are second and third order elastic tensor components, respectively. Data are from Ref. \cite{bogardus_thirdorder_1965,powell_combinations_1970,graham_determination_1972,yost_adiabatic_1973}.}\label{tab:1}
\end{table}

\begin{figure}[ht]
	\centering
	\includegraphics[width=7cm]{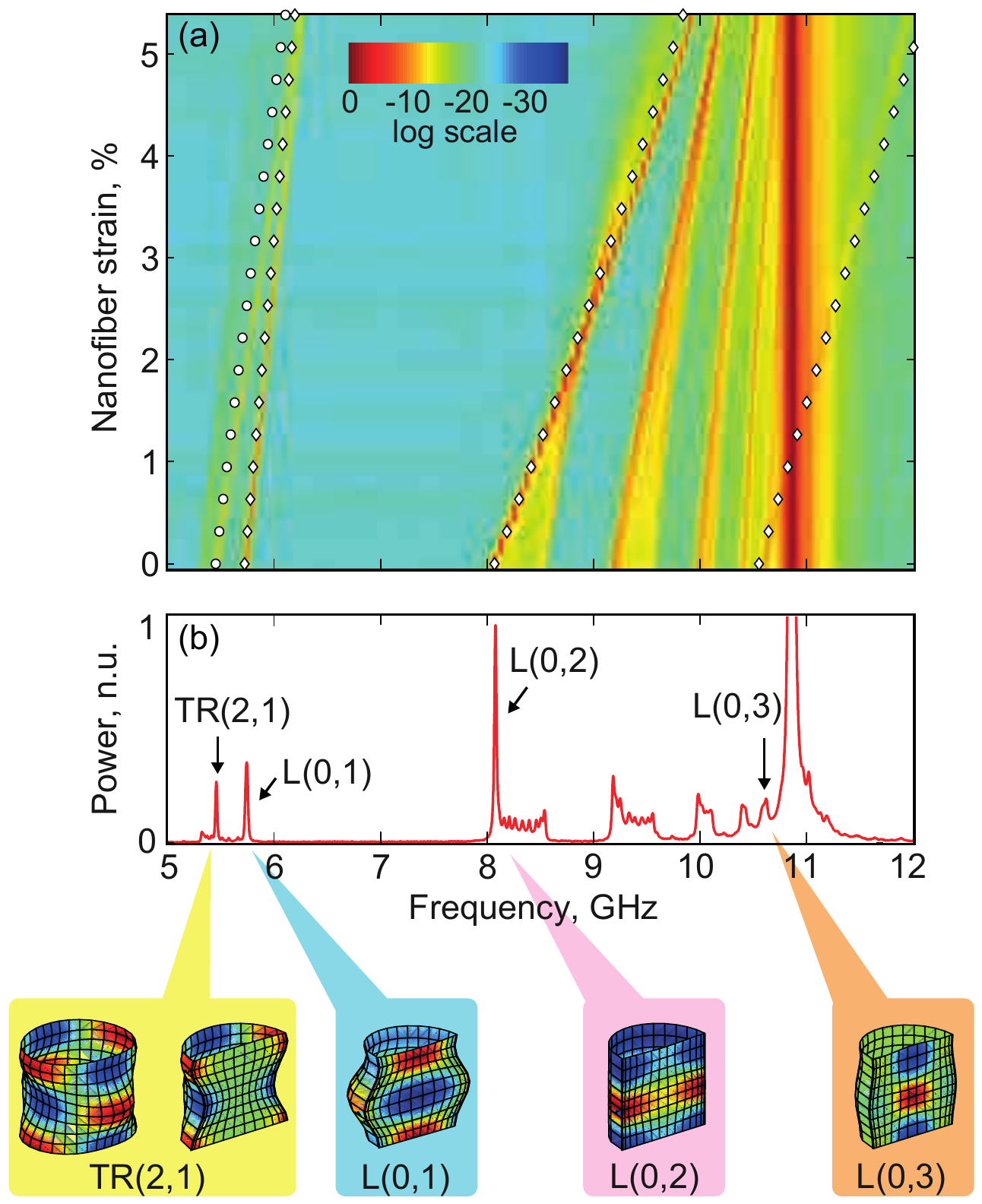} 
	\caption{(a) Comparison between our analytic model (white dots and diamonds) and the experimental Brillouin spectra for a nanofiber of diameter $d_{MF,1}$=660 nm versus tensile strain from 0 to $6\%$. Lower inset: Experimental Brillouin spectrum of the fiber without strain. (c) The insets show a simulation of both the transverse deformation and the longitudinal strain of the nanofiber for four acoustic modes including the TR(2,1), L(0,1), L(0,2) and L(0,3) , respectively. The vertical scale is normalized to roughly one acoustic wavelength.}
	\label{fig:Res1}
	\end{figure}
	
Computing the acoustic mode velocity in a silica rod with transverse isotropic symmetry is somewhat complicated \cite{honarvar_wave_2007}. In our case, however, the strong symmetry of our problem and the limited solutions we are interested in allow for simplifying the problem. For example, the longitudinal and shear waves approximation is still valid. The elastic coefficients related to shear waves are similarly affected by the acoustic nonlinearity, as shown by the lower-right submatrix coefficients, in which the pink and purple coefficients are quite close. Furthermore, although the upper-left submatrix diagonal coefficients are no longer the same, the phase-matching condition requires that the acoustic wave propagation occurs along the $z$ direction. Therefore the effect of the acoustic nonlinearity on $C'_{33}=v_z^2\rho=(\lambda+2\mu)\left(1+\frac{C_{111}}{\lambda + 2\mu}\bar{\epsilon}_{zz}\right)$ is one of major sources of perturbation, where $\rho$ is the material density. In Eq. \ref{eq:T}, the coefficient affecting the longitudinal velocity along $z$ is $\frac{C_{111}}{\lambda + 2\mu}=6.71$ (highlighted in blue). It increases the longitudinal velocity $v_z$ of a plane wave propagating along $z$ up to 20\% under high tensile strain. As a comparison, the longitudinal velocity for the other directions ($v_x$ and $v_y$) increases by 9\% only. The strong asymmetry of the elastic properties does not prevent to perform the standard computation of acoustic modes since the phase-matching condition for Brillouin backscattering is for waves propagating essentially along the $z$ direction. Given these observations, we assume that the displacement field is still the superposition of a pure longitudinal wave $\phi$ propagating along $z$ with the velocity $v_z=\sqrt{\frac{C'_{33}}{\rho}}$ and of a shear wave $\vec{\Psi}$ of velocity $v_s=\sqrt{\frac{C'_{44}}{\rho}}$, and therefore it can read as
\begin{equation}
\vec{u}= \vec{\nabla} \phi + \vec{\nabla} \wedge \vec{\psi},\label{eq:Assume}
\end{equation}where in cylindar coordinates,
\begin{eqnarray*}
\phi&=&A\mathrm{J_n}(pr)\cos (n\theta) \mathrm{e}^{\mathrm{i} (\Omega t-kz)},\\
\vec{\Psi}&=&
\left(
\begin{array}{c}
C\mathrm{J_{n+1}}(qr)\sin n\theta \\
-C\mathrm{J_{n+1}}(qr)\cos n\theta \\
B\mathrm{J_{n+1}}(qr)\sin n\theta \\
\end{array}\right)
\mathrm{e}^{\mathrm{i} (\Omega t-kz)},
\end{eqnarray*}
 with $p=\sqrt{\frac{\Omega^2}{v_z^2}-k^2}$ and $q=\sqrt{\frac{\Omega^2}{v_s^2}-k^2}$, n is the azimutal order, $\Omega$ the acoustic wave pulsation, k is the acoustic  longitudinal wavevector, A,B and C are amplitudes constants\cite{Royer_Elastic_2000}. From Eq. \ref{eq:02b} we then compute the stress component involved in the boundary conditions, and this yields for the Hooke's law
\begin{eqnarray}
	\sigma_{rr}&=& \mathcolorbox{LightY}{C'_{11}} \frac{\partial u_r}{\partial r}
	+\mathcolorbox{LightR}{C'_{12}}\left(
	\frac{1}{r}\frac{\partial u_\theta}{\partial \theta} + \frac{u_r}{r}
	\right)  
	+\mathcolorbox{LightG}{C'_{13}} \frac{\partial u_z}{\partial z}
	,\label{eq:RR}\\	
	\sigma_{r\theta}&=& \mathcolorbox{LightV}{C'_{66}} \left(\frac{1}{r}\frac{\partial u_r}{\partial \theta}+\frac{\partial u_\theta}{\partial r}\right),\\	
	\sigma_{rz}&=& \mathcolorbox{LightRo}{C'_{44}} \left( \frac{\partial u_r}{\partial z}+\frac{\partial u_z}{\partial r}\right).
\end{eqnarray}
The boundary conditions impose that these three components are zero at the surface of the nanofiber. This leads to a system of 3 equations with non-zero solutions if the determinant is null. Solving the resulting dispersion equation we find the acoustic modes wavevector $\beta_a$ from which we can get the Brillouin frequency shift. Comparing the coefficients highlighted in colors with their detailed expression in Eq. \ref{eq:T}, we find that the acoustic nonlinearity has a contribution about 3 times $\bar{\epsilon}_{zz}$ for most tensor coefficients, except one coefficient in the off diagonal green component $C'_{12}=\lambda\left(1+\frac{C_{112}}{\lambda}\bar{\epsilon}_{zz}\right)$ with $\frac{C_{112}}{\lambda}=15.05$. This coefficient alone has a big impact on the dispersion equation. As the boundary condition requires Eq. \ref{eq:RR} to be null, it strongly affects how the z-longitudinal strain component $\epsilon_{zz}=\frac{\partial u_z}{\partial z}$ is coupled to the transverse strains $\epsilon_{rr}=\frac{\partial u_r}{\partial r}$ and $\epsilon_{\theta\theta}=\frac{1}{r}\frac{\partial u_\theta}{\partial \theta} + \frac{u_r}{r}$.

\begin{figure}[ht]
	\centering
	\includegraphics[width=7cm]{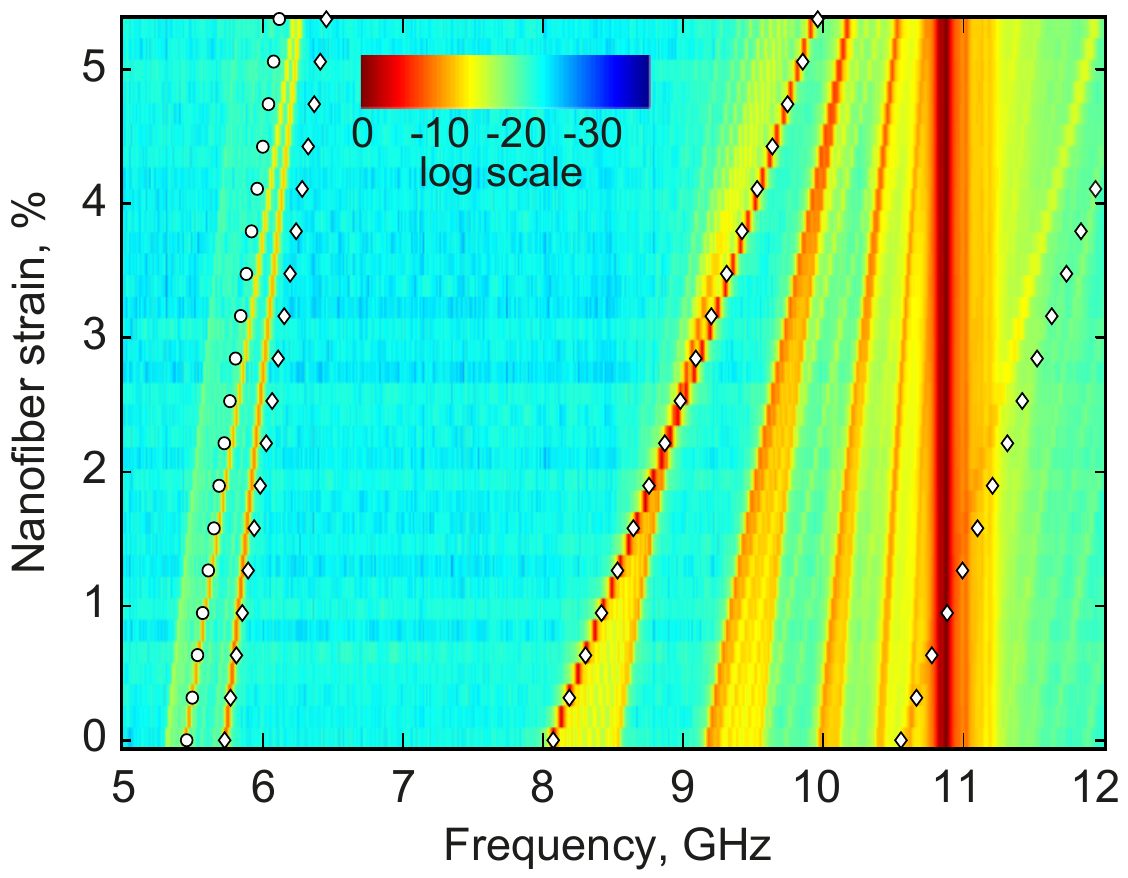} 
	\caption{Comparison between the analytic model with $C_{112}=0$ (white dots and diamonds) and the experimental Brillouin spectra for a nanofiber of diameter $d_{MF,1}$=660 nm versus tensile strain from 0 to $6\%$. When $C_{112}$ is zero, the anisotropy vanishes and the theory does not predict TR(2,1), L(0,1) Brillouin resonances crossing anymore.}
	\label{fig:Res1ANIS}
	\end{figure}
	
Once the dispersion equation is solved, we further need to combine the static and dynamic solutions to get the acoustic wavevector and to compute the Brillouin frequency shifts as a function of tensile strain applied on the nanofiber (see SI \ref{A2} for detailed calculations). Fig. \ref{fig:Res1}(a) shows the results of these calculations as white dots and diamonds superimposed on the experimental Brillouin spectra. The white dots describe the TR(2,1) Brillouin acoustic mode as a function of tensile strain while the white diamonds correspond to the L(0,x) longitudinal modes, with x=1-3. As can be seen, the agreement between theory and experiment is excellent. In particular, all the Brillouin frequency shifts and the crossing of TR(2,1) and L(0,1) modes are very well fitted by our model. 
To further illustrate the influence of the induced anisotropy, we show in Fig. \ref{fig:Res1ANIS}, the same comparison but without taking into account the asymmetry induced by $C'_{12}$ green component. Comparing Fig. \ref{fig:Res1}(a) and Fig. \ref{fig:Res1ANIS} shows that the anisotropy due to $C'_{12}$ is the main reason for the mode crossing between L(0,1) and TR(2,1). 
This mode crossing can be easily explained by looking at Fig. \ref{fig:Res1}(b) that shows the four main acoustic modes propagating in the nanofiber. Note that the deformation has been magnified for better visibility and that the color stands for the longitudinal deformation $\epsilon_{zz}$. When comparing the four modes in Fig. \ref{fig:Res1}(b), L(0,1) definitely shows the most hybrid nature. It has the strongest combination of both transverse deformation and longitudinal strain. In contrast, the TR(2,1) mode, highlighted in yellow, is almost a pure surface mode. L(0,2) and L(0,3), although being hybrids, are rather surface-like or longitudinal-like modes. The fact that $C'_{12}$ affects the coupling between the transverse deformations and the longitudinal strain explains why L(0,1) velocity is so strongly altered.

Although our analytic model provides a good agreement with experimental data for these four modes, it does not find the small frequency peak around 9 GHz in Fig. \ref{fig:Res1ANIS} and its weak slope under tensile strain. Once again, this remarkable behavior is in complete contradiction with the standard model described by Eq. \ref{eq:01} that scales with the Brillouin frequency. This is actually because these small peaks come from the two taper transitions and not from the nanofiber itself (see Fig. \ref{fig:Principe}). However, the two adiabatic taper transitions experience less tensile strain than the nanofiber and thus their contribution must be separately computed. To that end, we used a finite-element method (FEM), which allows for including the electrostrictive force and for modeling the exact solutions\cite{beugnotElectrostrictionGuidanceAcoustic2012a}. Attention must be paid however because the transitions have wider and varying diameters. They support a great number of acoustic modes, most of which are not excited because of a weak transverse overlap with the optical modes. By including the electrostrictive force in the simulations, we rule out most of the non excited acoustic modes.

	\begin{figure}[ht]
	\centering
\includegraphics[scale=0.5]{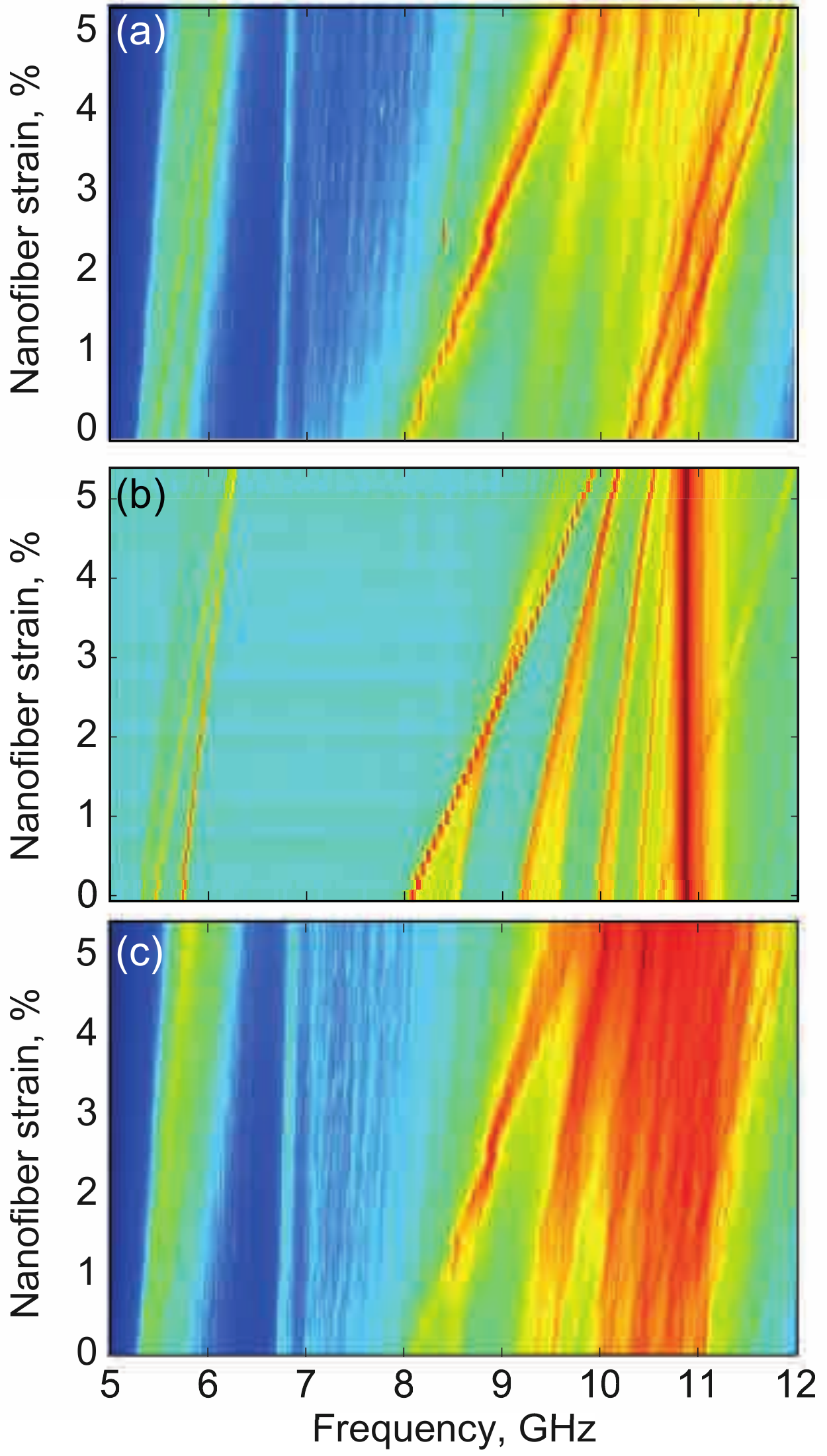}						
\caption{(a) Numerical simulations of the Brillouin spectrum including the uniform nanofiber, the taper transitions (up to 3 $\mu$m diameter), and the single-mode fiber.
(b) Experimental Brillouin spectra for a 660 nm-diameter tapered optical fiber versus strain.
(c) Numerical simulations of the tapered transitions only for diameter up to 3 $\mu$m. The tensile strain for each part of the transition was computed using Ref. \cite{Holleis_2014}.}
\label{fig:comsol}
\end{figure}
The numerical results are shown in Fig. \ref{fig:comsol}(a), it shows the Brillouin spectrum as a function of tensile strain for a nanofiber of 660 nm, including both the tapered and untapered fiber sections. As can be seen, the comparison with the experimental results shown in \ref{fig:comsol}(b) is very good and even better than with the analytical model. The computed Brillouin spectrum now clearly shows some peaks in the 9-10 GHz range at zero strain due to the tapered fiber transitions, as those experimentally observed. Unlike the analytic model, the FEM does not make any assumption on the solutions and allows for predicting the Brillouin spectrum of nanofibers for every diameter. To get further insight, Fig. \ref{fig:comsol}(c) shows the simulated Brillouin spectrum of the optical taper transitions only. One can clearly see the lines starting from 9 and 10 GHz without tensile strain, as experimentally shown in Fig. \ref{fig:comsol}(b).

\begin{figure}[ht]
	\centering
	\includegraphics[scale=0.4]{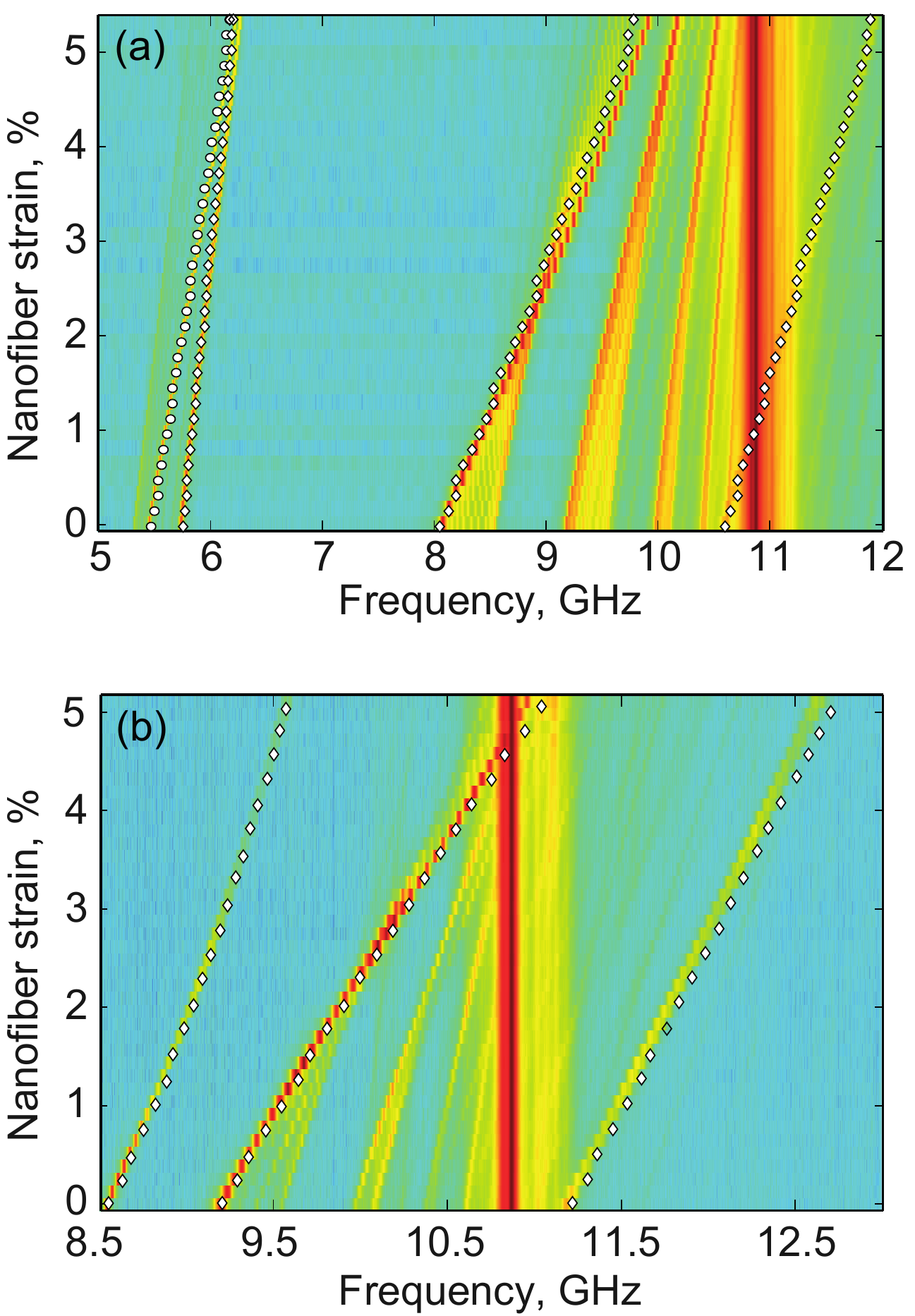}
					\caption{Comparison between the experimental Brillouin spectra in colormap vs finite-element method-based numerical simulations for two different diameter optical nanofibers : (a) 660 nm (b) 930 nm.}
\label{fig:diam}
\end{figure}

To confirm the model, comparison for different diameters was performed. In Figs. \ref{fig:diam}(a-c), we report the measurement and simulation data for 660 nm and 930 nm diameters. The Brillouin resonances were extracted from the simulation data by finding the local maxima of the simulated Brillouin spectrum. The superposition of experimental data and the simulated Brillouin resonances (dots) show a very good agreement for all diameters, as seen in Figs. \ref{fig:diam}(a-c). 
Furthermore, Fig. \ref{fig:diam}(c) shows that the Brillouin resonances for a 930 nm-diameter nanofiber nonlinearly shift with the tensile strain. This behavior is not due to the fourth-order elastic constant but by the complex acoustic dispersion. This nonlinear evolution is well predicted by both analytic and FEM models when TOECs are included. The nonlinear shift with respect to tensile strain is here due to the fact the two branches around 8 and 9  GHz are due to elastic waves strongly coupled at the nanofiber surface \cite{godet_brillouin_2017}. This strong coupling leads to an avoided crossing that dramatically affects the Brillouin resonances.

\subsection*{Conclusion}
To conclude, we have reported a detailed investigation of backward Brillouin scattering in tapered optical nanofibers under tensile strain. Our results revealed that the fundamental properties of acoustic waves dramatically change due to elastic anisotropy. We observed in particular unexpected and remarkable Brillouin strain coefficients, followed by a nonlinear evolution at high tensile strength. We further provided a complete theoretical model based on third-order nonlinear elasticity of silica that remarkably agrees with our experimental data. Our model is valid for a wide range of fiber taper diameters and it was further checked using FEM-based numerical simulations. The complex dispersion relation of acoustic modes is further responsible for the nonlinear behavior with respect to tensile strain. Finally, this work contributes to a further understanding of the complex light-sound interaction in optical waveguides and these new behaviors could open the way to the development of compact tensile strength optical sensors based on optical nanofibers. 
			
\section{Acknowledgement}
This work has received funding from the French National Research Agency under grant agreements ANR-16-CE24-0010-03, ANR-17-EURE-0002, and ANR-15-IDEX-0003. Adrien Godet thanks the Conseil R\'{e}gional de Bourgogne Franche-Comt\'{e} for student scholarship.

\bibliography{biblio4b}

\begin{thebibliography}{38}%
\makeatletter
\providecommand \@ifxundefined [1]{%
 \@ifx{#1\undefined}
}%
\providecommand \@ifnum [1]{%
 \ifnum #1\expandafter \@firstoftwo
 \else \expandafter \@secondoftwo
 \fi
}%
\providecommand \@ifx [1]{%
 \ifx #1\expandafter \@firstoftwo
 \else \expandafter \@secondoftwo
 \fi
}%
\providecommand \natexlab [1]{#1}%
\providecommand \enquote  [1]{``#1''}%
\providecommand \bibnamefont  [1]{#1}%
\providecommand \bibfnamefont [1]{#1}%
\providecommand \citenamefont [1]{#1}%
\providecommand \href@noop [0]{\@secondoftwo}%
\providecommand \href [0]{\begingroup \@sanitize@url \@href}%
\providecommand \@href[1]{\@@startlink{#1}\@@href}%
\providecommand \@@href[1]{\endgroup#1\@@endlink}%
\providecommand \@sanitize@url [0]{\catcode `\\12\catcode `\$12\catcode
  `\&12\catcode `\#12\catcode `\^12\catcode `\_12\catcode `\%12\relax}%
\providecommand \@@startlink[1]{}%
\providecommand \@@endlink[0]{}%
\providecommand \url  [0]{\begingroup\@sanitize@url \@url }%
\providecommand \@url [1]{\endgroup\@href {#1}{\urlprefix }}%
\providecommand \urlprefix  [0]{URL }%
\providecommand \Eprint [0]{\href }%
\providecommand \doibase [0]{http://dx.doi.org/}%
\providecommand \selectlanguage [0]{\@gobble}%
\providecommand \bibinfo  [0]{\@secondoftwo}%
\providecommand \bibfield  [0]{\@secondoftwo}%
\providecommand \translation [1]{[#1]}%
\providecommand \BibitemOpen [0]{}%
\providecommand \bibitemStop [0]{}%
\providecommand \bibitemNoStop [0]{.\EOS\space}%
\providecommand \EOS [0]{\spacefactor3000\relax}%
\providecommand \BibitemShut  [1]{\csname bibitem#1\endcsname}%
\let\auto@bib@innerbib\@empty
\bibitem [{\citenamefont {Birks}\ and\ \citenamefont
  {Li}(1992)}]{birks_shape_1992}%
  \BibitemOpen
  \bibfield  {author} {\bibinfo {author} {\bibfnamefont {T.~A.}\ \bibnamefont
  {Birks}}\ and\ \bibinfo {author} {\bibfnamefont {Y.~W.}\ \bibnamefont {Li}},\
  }\href {\doibase 10.1109/50.134196} {\bibfield  {journal} {\bibinfo
  {journal} {Journal of Lightwave Technology}\ }\textbf {\bibinfo {volume}
  {10}},\ \bibinfo {pages} {432} (\bibinfo {year} {1992})}\BibitemShut
  {NoStop}%
\bibitem [{\citenamefont {Tong}\ \emph {et~al.}(2003)\citenamefont {Tong},
  \citenamefont {Gattass}, \citenamefont {Ashcom}, \citenamefont {He},
  \citenamefont {Lou}, \citenamefont {Shen}, \citenamefont {Maxwell},\ and\
  \citenamefont {Mazur}}]{tong_subwavelength-diameter_2003}%
  \BibitemOpen
  \bibfield  {author} {\bibinfo {author} {\bibfnamefont {L.}~\bibnamefont
  {Tong}}, \bibinfo {author} {\bibfnamefont {R.~R.}\ \bibnamefont {Gattass}},
  \bibinfo {author} {\bibfnamefont {J.~B.}\ \bibnamefont {Ashcom}}, \bibinfo
  {author} {\bibfnamefont {S.}~\bibnamefont {He}}, \bibinfo {author}
  {\bibfnamefont {J.}~\bibnamefont {Lou}}, \bibinfo {author} {\bibfnamefont
  {M.}~\bibnamefont {Shen}}, \bibinfo {author} {\bibfnamefont {I.}~\bibnamefont
  {Maxwell}}, \ and\ \bibinfo {author} {\bibfnamefont {E.}~\bibnamefont
  {Mazur}},\ }\href {\doibase 10.1038/nature02193} {\bibfield  {journal}
  {\bibinfo  {journal} {Nature}\ }\textbf {\bibinfo {volume} {426}},\ \bibinfo
  {pages} {816} (\bibinfo {year} {2003})}\BibitemShut {NoStop}%
\bibitem [{\citenamefont {Tong}\ and\ \citenamefont
  {Sumetsky}(2010)}]{tong_book}%
  \BibitemOpen
  \bibfield  {author} {\bibinfo {author} {\bibfnamefont {L.}~\bibnamefont
  {Tong}}\ and\ \bibinfo {author} {\bibfnamefont {M.}~\bibnamefont
  {Sumetsky}},\ }\href@noop {} {\emph {\bibinfo {title} {Subwavelength and
  {{Nanometer Diameter Optical Fibers}}}}},\ Advanced Topics in Science and
  Technology in China\ (\bibinfo  {publisher} {{Springer-Verlag}},\ \bibinfo
  {address} {Berlin Heidelberg},\ \bibinfo {year} {2010})\BibitemShut {NoStop}%
\bibitem [{\citenamefont {Brambilla}(2010)}]{brambilla_optical_2010}%
  \BibitemOpen
  \bibfield  {author} {\bibinfo {author} {\bibfnamefont {G.}~\bibnamefont
  {Brambilla}},\ }\href {\doibase 10.1088/2040-8978/12/4/043001} {\bibfield
  {journal} {\bibinfo  {journal} {Journal of Optics}\ }\textbf {\bibinfo
  {volume} {12}},\ \bibinfo {pages} {043001} (\bibinfo {year}
  {2010})}\BibitemShut {NoStop}%
\bibitem [{\citenamefont {Foster}\ \emph {et~al.}(2008)\citenamefont {Foster},
  \citenamefont {Turner}, \citenamefont {Lipson},\ and\ \citenamefont
  {Gaeta}}]{foster_nonlinear_2008}%
  \BibitemOpen
  \bibfield  {author} {\bibinfo {author} {\bibfnamefont {M.~A.}\ \bibnamefont
  {Foster}}, \bibinfo {author} {\bibfnamefont {A.~C.}\ \bibnamefont {Turner}},
  \bibinfo {author} {\bibfnamefont {M.}~\bibnamefont {Lipson}}, \ and\ \bibinfo
  {author} {\bibfnamefont {A.~L.}\ \bibnamefont {Gaeta}},\ }\href {\doibase
  10.1364/OE.16.001300} {\bibfield  {journal} {\bibinfo  {journal} {Optics
  Express}\ }\textbf {\bibinfo {volume} {16}},\ \bibinfo {pages} {1300}
  (\bibinfo {year} {2008})}\BibitemShut {NoStop}%
\bibitem [{\citenamefont {Sayrin}\ \emph {et~al.}(2015)\citenamefont {Sayrin},
  \citenamefont {Clausen}, \citenamefont {Albrecht}, \citenamefont
  {Schneeweiss},\ and\ \citenamefont {Rauschenbeutel}}]{sayrin_storage_2015}%
  \BibitemOpen
  \bibfield  {author} {\bibinfo {author} {\bibfnamefont {C.}~\bibnamefont
  {Sayrin}}, \bibinfo {author} {\bibfnamefont {C.}~\bibnamefont {Clausen}},
  \bibinfo {author} {\bibfnamefont {B.}~\bibnamefont {Albrecht}}, \bibinfo
  {author} {\bibfnamefont {P.}~\bibnamefont {Schneeweiss}}, \ and\ \bibinfo
  {author} {\bibfnamefont {A.}~\bibnamefont {Rauschenbeutel}},\ }\href
  {\doibase 10.1364/OPTICA.2.000353} {\bibfield  {journal} {\bibinfo  {journal}
  {Optica}\ }\textbf {\bibinfo {volume} {2}},\ \bibinfo {pages} {353} (\bibinfo
  {year} {2015})}\BibitemShut {NoStop}%
\bibitem [{\citenamefont {Gouraud}\ \emph {et~al.}(2015)\citenamefont
  {Gouraud}, \citenamefont {Maxein}, \citenamefont {Nicolas}, \citenamefont
  {Morin},\ and\ \citenamefont {Laurat}}]{gouraud_demonstration_2015}%
  \BibitemOpen
  \bibfield  {author} {\bibinfo {author} {\bibfnamefont {B.}~\bibnamefont
  {Gouraud}}, \bibinfo {author} {\bibfnamefont {D.}~\bibnamefont {Maxein}},
  \bibinfo {author} {\bibfnamefont {A.}~\bibnamefont {Nicolas}}, \bibinfo
  {author} {\bibfnamefont {O.}~\bibnamefont {Morin}}, \ and\ \bibinfo {author}
  {\bibfnamefont {J.}~\bibnamefont {Laurat}},\ }\href {\doibase
  10.1103/PhysRevLett.114.180503} {\bibfield  {journal} {\bibinfo  {journal}
  {Physical Review Letters}\ }\textbf {\bibinfo {volume} {114}},\ \bibinfo
  {pages} {180503} (\bibinfo {year} {2015})}\BibitemShut {NoStop}%
\bibitem [{\citenamefont {Holleis}\ \emph {et~al.}(2014)\citenamefont
  {Holleis}, \citenamefont {Hoinkes}, \citenamefont {Wuttke}, \citenamefont
  {Schneeweiss},\ and\ \citenamefont {Rauschenbeutel}}]{Holleis_2014}%
  \BibitemOpen
  \bibfield  {author} {\bibinfo {author} {\bibfnamefont {S.}~\bibnamefont
  {Holleis}}, \bibinfo {author} {\bibfnamefont {T.}~\bibnamefont {Hoinkes}},
  \bibinfo {author} {\bibfnamefont {C.}~\bibnamefont {Wuttke}}, \bibinfo
  {author} {\bibfnamefont {P.}~\bibnamefont {Schneeweiss}}, \ and\ \bibinfo
  {author} {\bibfnamefont {A.}~\bibnamefont {Rauschenbeutel}},\ }\href
  {\doibase 10.1063/1.4873339} {\bibfield  {journal} {\bibinfo  {journal}
  {Applied Physics Letters}\ }\textbf {\bibinfo {volume} {104}},\ \bibinfo
  {pages} {163109} (\bibinfo {year} {2014})}\BibitemShut {NoStop}%
\bibitem [{\citenamefont {Beugnot}\ \emph {et~al.}(2014)\citenamefont
  {Beugnot}, \citenamefont {Lebrun}, \citenamefont {Pauliat}, \citenamefont
  {Maillotte}, \citenamefont {Laude},\ and\ \citenamefont
  {Sylvestre}}]{beugnot_brillouin_2014}%
  \BibitemOpen
  \bibfield  {author} {\bibinfo {author} {\bibfnamefont {J.-C.}\ \bibnamefont
  {Beugnot}}, \bibinfo {author} {\bibfnamefont {S.}~\bibnamefont {Lebrun}},
  \bibinfo {author} {\bibfnamefont {G.}~\bibnamefont {Pauliat}}, \bibinfo
  {author} {\bibfnamefont {H.}~\bibnamefont {Maillotte}}, \bibinfo {author}
  {\bibfnamefont {V.}~\bibnamefont {Laude}}, \ and\ \bibinfo {author}
  {\bibfnamefont {T.}~\bibnamefont {Sylvestre}},\ }\href {\doibase
  10.1038/ncomms6242} {\bibfield  {journal} {\bibinfo  {journal} {Nature
  Communications}\ }\textbf {\bibinfo {volume} {5}},\ \bibinfo {pages} {5242}
  (\bibinfo {year} {2014})}\BibitemShut {NoStop}%
\bibitem [{\citenamefont {Florez}\ \emph {et~al.}(2016)\citenamefont {Florez},
  \citenamefont {Jarschel}, \citenamefont {Espinel}, \citenamefont {Cordeiro},
  \citenamefont {Alegre}, \citenamefont {Wiederhecker},\ and\ \citenamefont
  {Dainese}}]{florez_brillouin_2016}%
  \BibitemOpen
  \bibfield  {author} {\bibinfo {author} {\bibfnamefont {O.}~\bibnamefont
  {Florez}}, \bibinfo {author} {\bibfnamefont {P.~F.}\ \bibnamefont
  {Jarschel}}, \bibinfo {author} {\bibfnamefont {Y.~a.~V.}\ \bibnamefont
  {Espinel}}, \bibinfo {author} {\bibfnamefont {C.~M.~B.}\ \bibnamefont
  {Cordeiro}}, \bibinfo {author} {\bibfnamefont {T.~P.~M.}\ \bibnamefont
  {Alegre}}, \bibinfo {author} {\bibfnamefont {G.~S.}\ \bibnamefont
  {Wiederhecker}}, \ and\ \bibinfo {author} {\bibfnamefont {P.}~\bibnamefont
  {Dainese}},\ }\href {\doibase 10.1038/ncomms11759} {\bibfield  {journal}
  {\bibinfo  {journal} {Nature Communications}\ }\textbf {\bibinfo {volume}
  {7}},\ \bibinfo {pages} {11759} (\bibinfo {year} {2016})}\BibitemShut
  {NoStop}%
\bibitem [{\citenamefont {Brillouin}(1922)}]{brillouin_diffusion_1922}%
  \BibitemOpen
  \bibfield  {author} {\bibinfo {author} {\bibfnamefont {L.}~\bibnamefont
  {Brillouin}},\ }\href {\doibase 10.1051/anphys/192209170088} {\bibfield
  {journal} {\bibinfo  {journal} {Annales de Physique}\ }\textbf {\bibinfo
  {volume} {9}},\ \bibinfo {pages} {88} (\bibinfo {year} {1922})}\BibitemShut
  {NoStop}%
\bibitem [{\citenamefont {Brillouin}(1925)}]{brillouin_les_1925}%
  \BibitemOpen
  \bibfield  {author} {\bibinfo {author} {\bibfnamefont {L.}~\bibnamefont
  {Brillouin}},\ }\href {\doibase 10.1051/anphys/192510030251} {\bibfield
  {journal} {\bibinfo  {journal} {Annales de Physique}\ }\textbf {\bibinfo
  {volume} {10}},\ \bibinfo {pages} {251} (\bibinfo {year} {1925})}\BibitemShut
  {NoStop}%
\bibitem [{\citenamefont {Brillouin}(1946)}]{brillouin_les_1946}%
  \BibitemOpen
  \bibfield  {author} {\bibinfo {author} {\bibfnamefont {L.}~\bibnamefont
  {Brillouin}},\ }\href@noop {} {\emph {\bibinfo {title} {Les tenseurs en
  m\'ecanique et en \'elasticit\'e}}}\ (\bibinfo  {publisher} {Dover
  Publications},\ \bibinfo {year} {1946})\BibitemShut {NoStop}%
\bibitem [{\citenamefont {Hughes}\ and\ \citenamefont
  {Kelly}(1953)}]{hughes_second-order_1953}%
  \BibitemOpen
  \bibfield  {author} {\bibinfo {author} {\bibfnamefont {D.~S.}\ \bibnamefont
  {Hughes}}\ and\ \bibinfo {author} {\bibfnamefont {J.~L.}\ \bibnamefont
  {Kelly}},\ }\href {\doibase 10.1103/PhysRev.92.1145} {\bibfield  {journal}
  {\bibinfo  {journal} {Physical Review}\ }\textbf {\bibinfo {volume} {92}},\
  \bibinfo {pages} {1145} (\bibinfo {year} {1953})}\BibitemShut {NoStop}%
\bibitem [{\citenamefont {Bogardus}(1965)}]{bogardus_thirdorder_1965}%
  \BibitemOpen
  \bibfield  {author} {\bibinfo {author} {\bibfnamefont {E.~H.}\ \bibnamefont
  {Bogardus}},\ }\href {\doibase 10.1063/1.1714520} {\bibfield  {journal}
  {\bibinfo  {journal} {Journal of Applied Physics}\ }\textbf {\bibinfo
  {volume} {36}},\ \bibinfo {pages} {2504} (\bibinfo {year}
  {1965})}\BibitemShut {NoStop}%
\bibitem [{\citenamefont {Wang}\ \emph {et~al.}(1992)\citenamefont {Wang},
  \citenamefont {Saunders}, \citenamefont {Senin},\ and\ \citenamefont
  {Lambson}}]{Wang_temperature_1992}%
  \BibitemOpen
  \bibfield  {author} {\bibinfo {author} {\bibfnamefont {Q.}~\bibnamefont
  {Wang}}, \bibinfo {author} {\bibfnamefont {G.~A.}\ \bibnamefont {Saunders}},
  \bibinfo {author} {\bibfnamefont {H.~B.}\ \bibnamefont {Senin}}, \ and\
  \bibinfo {author} {\bibfnamefont {E.~F.}\ \bibnamefont {Lambson}},\ }\href
  {\doibase 10.1016/S0022-3093(05)80554-X} {\bibfield  {journal} {\bibinfo
  {journal} {Journal of Non-Crystalline Solids}\ }\textbf {\bibinfo {volume}
  {143}},\ \bibinfo {pages} {65} (\bibinfo {year} {1992})}\BibitemShut
  {NoStop}%
\bibitem [{\citenamefont {Pine}(1969)}]{pine_brillouin_1969}%
  \BibitemOpen
  \bibfield  {author} {\bibinfo {author} {\bibfnamefont {A.~S.}\ \bibnamefont
  {Pine}},\ }\href {\doibase 10.1103/PhysRev.185.1187} {\bibfield  {journal}
  {\bibinfo  {journal} {Physical Review}\ }\textbf {\bibinfo {volume} {185}},\
  \bibinfo {pages} {1187} (\bibinfo {year} {1969})}\BibitemShut {NoStop}%
\bibitem [{\citenamefont {Vacher}\ and\ \citenamefont
  {Pelous}(1975)}]{vacher_temperature_1975}%
  \BibitemOpen
  \bibfield  {author} {\bibinfo {author} {\bibnamefont {Vacher}}\ and\ \bibinfo
  {author} {\bibnamefont {Pelous}},\ }\href {\doibase
  10.1016/0375-9601(75)90420-X} {\bibfield  {journal} {\bibinfo  {journal}
  {Physics Letters A}\ }\textbf {\bibinfo {volume} {53}},\ \bibinfo {pages}
  {233} (\bibinfo {year} {1975})}\BibitemShut {NoStop}%
\bibitem [{\citenamefont {Pelous}\ and\ \citenamefont
  {Vacher}(1976)}]{pelous_thermal_1976}%
  \BibitemOpen
  \bibfield  {author} {\bibinfo {author} {\bibnamefont {Pelous}}\ and\ \bibinfo
  {author} {\bibnamefont {Vacher}},\ }\href {\doibase
  10.1016/0038-1098(76)91505-2} {\bibfield  {journal} {\bibinfo  {journal}
  {Solid State Communications}\ }\textbf {\bibinfo {volume} {18}},\ \bibinfo
  {pages} {657} (\bibinfo {year} {1976})}\BibitemShut {NoStop}%
\bibitem [{\citenamefont {Horiguchi}, \citenamefont {Kurashima},\ and\
  \citenamefont {Tateda}(1989)}]{horiguchi_tensile_1989}%
  \BibitemOpen
  \bibfield  {author} {\bibinfo {author} {\bibfnamefont {T.}~\bibnamefont
  {Horiguchi}}, \bibinfo {author} {\bibfnamefont {T.}~\bibnamefont
  {Kurashima}}, \ and\ \bibinfo {author} {\bibfnamefont {M.}~\bibnamefont
  {Tateda}},\ }\href {\doibase 10.1109/68.34756} {\bibfield  {journal}
  {\bibinfo  {journal} {IEEE Photonics Technology Letters}\ }\textbf {\bibinfo
  {volume} {1}},\ \bibinfo {pages} {107} (\bibinfo {year} {1989})}\BibitemShut
  {NoStop}%
\bibitem [{\citenamefont {{Galindez-Jamioy}}\ and\ \citenamefont
  {{L\'opez-Higuera}}(2012)}]{Galindez_2012}%
  \BibitemOpen
  \bibfield  {author} {\bibinfo {author} {\bibfnamefont {C.~A.}\ \bibnamefont
  {{Galindez-Jamioy}}}\ and\ \bibinfo {author} {\bibfnamefont {J.~M.}\
  \bibnamefont {{L\'opez-Higuera}}},\ }\href {\doibase 10.1155/2012/204121}
  {\enquote {\bibinfo {title} {Brillouin {{Distributed Fiber Sensors}}: {{An
  Overview}} and {{Applications}}},}\ }\bibinfo {howpublished}
  {https://www.hindawi.com/journals/js/2012/204121/abs/} (\bibinfo {year}
  {2012})\BibitemShut {NoStop}%
\bibitem [{\citenamefont {Culverhouse}\ \emph {et~al.}(1989)\citenamefont
  {Culverhouse}, \citenamefont {Farahi}, \citenamefont {Pannell},\ and\
  \citenamefont {Jackson}}]{Culverhouse_1989}%
  \BibitemOpen
  \bibfield  {author} {\bibinfo {author} {\bibfnamefont {D.}~\bibnamefont
  {Culverhouse}}, \bibinfo {author} {\bibfnamefont {F.}~\bibnamefont {Farahi}},
  \bibinfo {author} {\bibfnamefont {C.~N.}\ \bibnamefont {Pannell}}, \ and\
  \bibinfo {author} {\bibfnamefont {D.~A.}\ \bibnamefont {Jackson}},\ }\href
  {\doibase 10.1049/el:19890612} {\bibfield  {journal} {\bibinfo  {journal}
  {Electronics Letters}\ }\textbf {\bibinfo {volume} {25}},\ \bibinfo {pages}
  {913} (\bibinfo {year} {1989})}\BibitemShut {NoStop}%
\bibitem [{\citenamefont {Horiguchi}, \citenamefont {Kurashima},\ and\
  \citenamefont {Tateda}(1990)}]{horiguchi_1990}%
  \BibitemOpen
  \bibfield  {author} {\bibinfo {author} {\bibfnamefont {T.}~\bibnamefont
  {Horiguchi}}, \bibinfo {author} {\bibfnamefont {T.}~\bibnamefont
  {Kurashima}}, \ and\ \bibinfo {author} {\bibfnamefont {M.}~\bibnamefont
  {Tateda}},\ }\href {\doibase 10.1109/68.54703} {\bibfield  {journal}
  {\bibinfo  {journal} {IEEE Photonics Technology Letters}\ }\textbf {\bibinfo
  {volume} {2}},\ \bibinfo {pages} {352} (\bibinfo {year} {1990})}\BibitemShut
  {NoStop}%
\bibitem [{\citenamefont {Nikl\`es}, \citenamefont {Th\'evenaz},\ and\
  \citenamefont {Robert}(1996)}]{Nikles_1996}%
  \BibitemOpen
  \bibfield  {author} {\bibinfo {author} {\bibfnamefont {M.}~\bibnamefont
  {Nikl\`es}}, \bibinfo {author} {\bibfnamefont {L.}~\bibnamefont
  {Th\'evenaz}}, \ and\ \bibinfo {author} {\bibfnamefont {P.~A.}\ \bibnamefont
  {Robert}},\ }\href {\doibase 10.1364/OL.21.000758} {\bibfield  {journal}
  {\bibinfo  {journal} {Optics Letters}\ }\textbf {\bibinfo {volume} {21}},\
  \bibinfo {pages} {758} (\bibinfo {year} {1996})}\BibitemShut {NoStop}%
\bibitem [{\citenamefont {Guerette}\ \emph {et~al.}(2016)\citenamefont
  {Guerette}, \citenamefont {Kurkjian}, \citenamefont {Semjonov},\ and\
  \citenamefont {Huang}}]{gueretteNonlinearElasticitySilica2016}%
  \BibitemOpen
  \bibfield  {author} {\bibinfo {author} {\bibfnamefont {M.}~\bibnamefont
  {Guerette}}, \bibinfo {author} {\bibfnamefont {C.~R.}\ \bibnamefont
  {Kurkjian}}, \bibinfo {author} {\bibfnamefont {S.}~\bibnamefont {Semjonov}},
  \ and\ \bibinfo {author} {\bibfnamefont {L.}~\bibnamefont {Huang}},\ }\href
  {\doibase 10.1111/jace.14043} {\bibfield  {journal} {\bibinfo  {journal}
  {Journal of the American Ceramic Society}\ }\textbf {\bibinfo {volume}
  {99}},\ \bibinfo {pages} {841} (\bibinfo {year} {2016})}\BibitemShut
  {NoStop}%
\bibitem [{\citenamefont {Godet}\ \emph {et~al.}(2017)\citenamefont {Godet},
  \citenamefont {Ndao}, \citenamefont {Sylvestre}, \citenamefont {Pecheur},
  \citenamefont {Lebrun}, \citenamefont {Pauliat}, \citenamefont {Beugnot},\
  and\ \citenamefont {Huy}}]{godet_brillouin_2017}%
  \BibitemOpen
  \bibfield  {author} {\bibinfo {author} {\bibfnamefont {A.}~\bibnamefont
  {Godet}}, \bibinfo {author} {\bibfnamefont {A.}~\bibnamefont {Ndao}},
  \bibinfo {author} {\bibfnamefont {T.}~\bibnamefont {Sylvestre}}, \bibinfo
  {author} {\bibfnamefont {V.}~\bibnamefont {Pecheur}}, \bibinfo {author}
  {\bibfnamefont {S.}~\bibnamefont {Lebrun}}, \bibinfo {author} {\bibfnamefont
  {G.}~\bibnamefont {Pauliat}}, \bibinfo {author} {\bibfnamefont {J.-C.}\
  \bibnamefont {Beugnot}}, \ and\ \bibinfo {author} {\bibfnamefont {K.~P.}\
  \bibnamefont {Huy}},\ }\href {\doibase 10.1364/OPTICA.4.001232} {\bibfield
  {journal} {\bibinfo  {journal} {Optica}\ }\textbf {\bibinfo {volume} {4}},\
  \bibinfo {pages} {1232} (\bibinfo {year} {2017})}\BibitemShut {NoStop}%
\bibitem [{\citenamefont {Rowell}\ and\ \citenamefont
  {Stegeman}(1978)}]{rowell_brillouin_1978}%
  \BibitemOpen
  \bibfield  {author} {\bibinfo {author} {\bibfnamefont {N.~L.}\ \bibnamefont
  {Rowell}}\ and\ \bibinfo {author} {\bibfnamefont {G.~I.}\ \bibnamefont
  {Stegeman}},\ }\href {\doibase 10.1103/PhysRevLett.41.970} {\bibfield
  {journal} {\bibinfo  {journal} {Physical Review Letters}\ }\textbf {\bibinfo
  {volume} {41}},\ \bibinfo {pages} {970} (\bibinfo {year} {1978})}\BibitemShut
  {NoStop}%
\bibitem [{\citenamefont {Volkov}\ \emph {et~al.}(2015)\citenamefont {Volkov},
  \citenamefont {Kokshaiskii}, \citenamefont {Korobov},\ and\ \citenamefont
  {Prokhorov}}]{volkov_second-_2015}%
  \BibitemOpen
  \bibfield  {author} {\bibinfo {author} {\bibfnamefont {A.~D.}\ \bibnamefont
  {Volkov}}, \bibinfo {author} {\bibfnamefont {A.~I.}\ \bibnamefont
  {Kokshaiskii}}, \bibinfo {author} {\bibfnamefont {A.~I.}\ \bibnamefont
  {Korobov}}, \ and\ \bibinfo {author} {\bibfnamefont {V.~M.}\ \bibnamefont
  {Prokhorov}},\ }\href {\doibase 10.1134/S1063771015060147} {\bibfield
  {journal} {\bibinfo  {journal} {Acoustical Physics}\ }\textbf {\bibinfo
  {volume} {61}},\ \bibinfo {pages} {651} (\bibinfo {year} {2015})}\BibitemShut
  {NoStop}%
\bibitem [{\citenamefont {Tanaka}\ and\ \citenamefont
  {Ogusu}(1999)}]{Tanaka_1999}%
  \BibitemOpen
  \bibfield  {author} {\bibinfo {author} {\bibfnamefont {Y.}~\bibnamefont
  {Tanaka}}\ and\ \bibinfo {author} {\bibfnamefont {K.}~\bibnamefont {Ogusu}},\
  }\href {\doibase 10.1109/68.769734} {\bibfield  {journal} {\bibinfo
  {journal} {IEEE Photonics Technology Letters}\ }\textbf {\bibinfo {volume}
  {11}},\ \bibinfo {pages} {865} (\bibinfo {year} {1999})}\BibitemShut
  {NoStop}%
\bibitem [{\citenamefont {Kobyakov}, \citenamefont {Sauer},\ and\ \citenamefont
  {Chowdhury}(2010)}]{kobyakov_stimulated_2010}%
  \BibitemOpen
  \bibfield  {author} {\bibinfo {author} {\bibfnamefont {A.}~\bibnamefont
  {Kobyakov}}, \bibinfo {author} {\bibfnamefont {M.}~\bibnamefont {Sauer}}, \
  and\ \bibinfo {author} {\bibfnamefont {D.}~\bibnamefont {Chowdhury}},\ }\href
  {\doibase 10.1364/AOP.2.000001} {\bibfield  {journal} {\bibinfo  {journal}
  {Advances in Optics and Photonics}\ }\textbf {\bibinfo {volume} {2}},\
  \bibinfo {pages} {1} (\bibinfo {year} {2010})}\BibitemShut {NoStop}%
\bibitem [{\citenamefont {Boyd}(2008)}]{boydNonlinearOpticsThird2008}%
  \BibitemOpen
  \bibfield  {author} {\bibinfo {author} {\bibfnamefont {R.~W.}\ \bibnamefont
  {Boyd}},\ }\href@noop {} {\emph {\bibinfo {title} {Nonlinear {{Optics}},
  {{Third Edition}}}}},\ \bibinfo {edition} {3rd}\ ed.\ (\bibinfo  {publisher}
  {{Academic Press}},\ \bibinfo {address} {Amsterdam ; Boston},\ \bibinfo
  {year} {2008})\BibitemShut {NoStop}%
\bibitem [{\citenamefont {Suhir}(1993)}]{Suhir_predicted_1993}%
  \BibitemOpen
  \bibfield  {author} {\bibinfo {author} {\bibfnamefont {E.}~\bibnamefont
  {Suhir}},\ }\href {\doibase 10.1364/AO.32.003237} {\bibfield  {journal}
  {\bibinfo  {journal} {Applied Optics}\ }\textbf {\bibinfo {volume} {32}},\
  \bibinfo {pages} {3237} (\bibinfo {year} {1993})}\BibitemShut {NoStop}%
\bibitem [{\citenamefont {Powell}\ and\ \citenamefont
  {Skove}(1970)}]{powell_combinations_1970}%
  \BibitemOpen
  \bibfield  {author} {\bibinfo {author} {\bibfnamefont {B.~E.}\ \bibnamefont
  {Powell}}\ and\ \bibinfo {author} {\bibfnamefont {M.~J.}\ \bibnamefont
  {Skove}},\ }\href {\doibase 10.1063/1.1658561} {\bibfield  {journal}
  {\bibinfo  {journal} {Journal of Applied Physics}\ }\textbf {\bibinfo
  {volume} {41}},\ \bibinfo {pages} {4913} (\bibinfo {year}
  {1970})}\BibitemShut {NoStop}%
\bibitem [{\citenamefont {Graham}(1972)}]{graham_determination_1972}%
  \BibitemOpen
  \bibfield  {author} {\bibinfo {author} {\bibfnamefont {R.~A.}\ \bibnamefont
  {Graham}},\ }\href {\doibase 10.1121/1.1913001} {\bibfield  {journal}
  {\bibinfo  {journal} {The Journal of the Acoustical Society of America}\
  }\textbf {\bibinfo {volume} {51}},\ \bibinfo {pages} {1576} (\bibinfo {year}
  {1972})}\BibitemShut {NoStop}%
\bibitem [{\citenamefont {Yost}\ and\ \citenamefont
  {Breazeale}(1973)}]{yost_adiabatic_1973}%
  \BibitemOpen
  \bibfield  {author} {\bibinfo {author} {\bibfnamefont {W.~T.}\ \bibnamefont
  {Yost}}\ and\ \bibinfo {author} {\bibfnamefont {M.~A.}\ \bibnamefont
  {Breazeale}},\ }\href {\doibase 10.1063/1.1662477} {\bibfield  {journal}
  {\bibinfo  {journal} {Journal of Applied Physics}\ }\textbf {\bibinfo
  {volume} {44}},\ \bibinfo {pages} {1909} (\bibinfo {year}
  {1973})}\BibitemShut {NoStop}%
\bibitem [{\citenamefont {Honarvar}\ \emph {et~al.}(2007)\citenamefont
  {Honarvar}, \citenamefont {Enjilela}, \citenamefont {Sinclair},\ and\
  \citenamefont {Mirnezami}}]{honarvar_wave_2007}%
  \BibitemOpen
  \bibfield  {author} {\bibinfo {author} {\bibfnamefont {F.}~\bibnamefont
  {Honarvar}}, \bibinfo {author} {\bibfnamefont {E.}~\bibnamefont {Enjilela}},
  \bibinfo {author} {\bibfnamefont {A.~N.}\ \bibnamefont {Sinclair}}, \ and\
  \bibinfo {author} {\bibfnamefont {S.~A.}\ \bibnamefont {Mirnezami}},\ }\href
  {\doibase 10.1016/j.ijsolstr.2006.12.029} {\bibfield  {journal} {\bibinfo
  {journal} {International Journal of Solids and Structures}\ }\textbf
  {\bibinfo {volume} {16}},\ \bibinfo {pages} {5236} (\bibinfo {year}
  {2007})}\BibitemShut {NoStop}%
\bibitem [{\citenamefont {Royer}\ and\ \citenamefont
  {Dieulesaint}(2000)}]{Royer_Elastic_2000}%
  \BibitemOpen
  \bibfield  {author} {\bibinfo {author} {\bibfnamefont {D.}~\bibnamefont
  {Royer}}\ and\ \bibinfo {author} {\bibfnamefont {E.}~\bibnamefont
  {Dieulesaint}},\ }\href@noop {} {{\selectlanguage {english}\emph {\bibinfo
  {title} {Elastic Waves in Solids I: Free and Guided Propagation}}}}\
  (\bibinfo  {publisher} {{Springer}},\ \bibinfo {year} {2000})\BibitemShut
  {NoStop}%
\bibitem [{\citenamefont {Beugnot}\ and\ \citenamefont
  {Laude}(2012)}]{beugnotElectrostrictionGuidanceAcoustic2012a}%
  \BibitemOpen
  \bibfield  {author} {\bibinfo {author} {\bibfnamefont {J.-C.}\ \bibnamefont
  {Beugnot}}\ and\ \bibinfo {author} {\bibfnamefont {V.}~\bibnamefont
  {Laude}},\ }\href {\doibase 10.1103/PhysRevB.86.224304} {\bibfield  {journal}
  {\bibinfo  {journal} {Physical Review B}\ }\textbf {\bibinfo {volume} {86}},\
  \bibinfo {pages} {224304} (\bibinfo {year} {2012})}\BibitemShut {NoStop}%
\end{thebibliography}%

\section*{Supporting information}
\setcounter{section}{0}
\section{Calculation of the nanofiber static strain}\label{A1}
\begin{figure}[ht]
	\centering
            \includegraphics[scale=0.5]{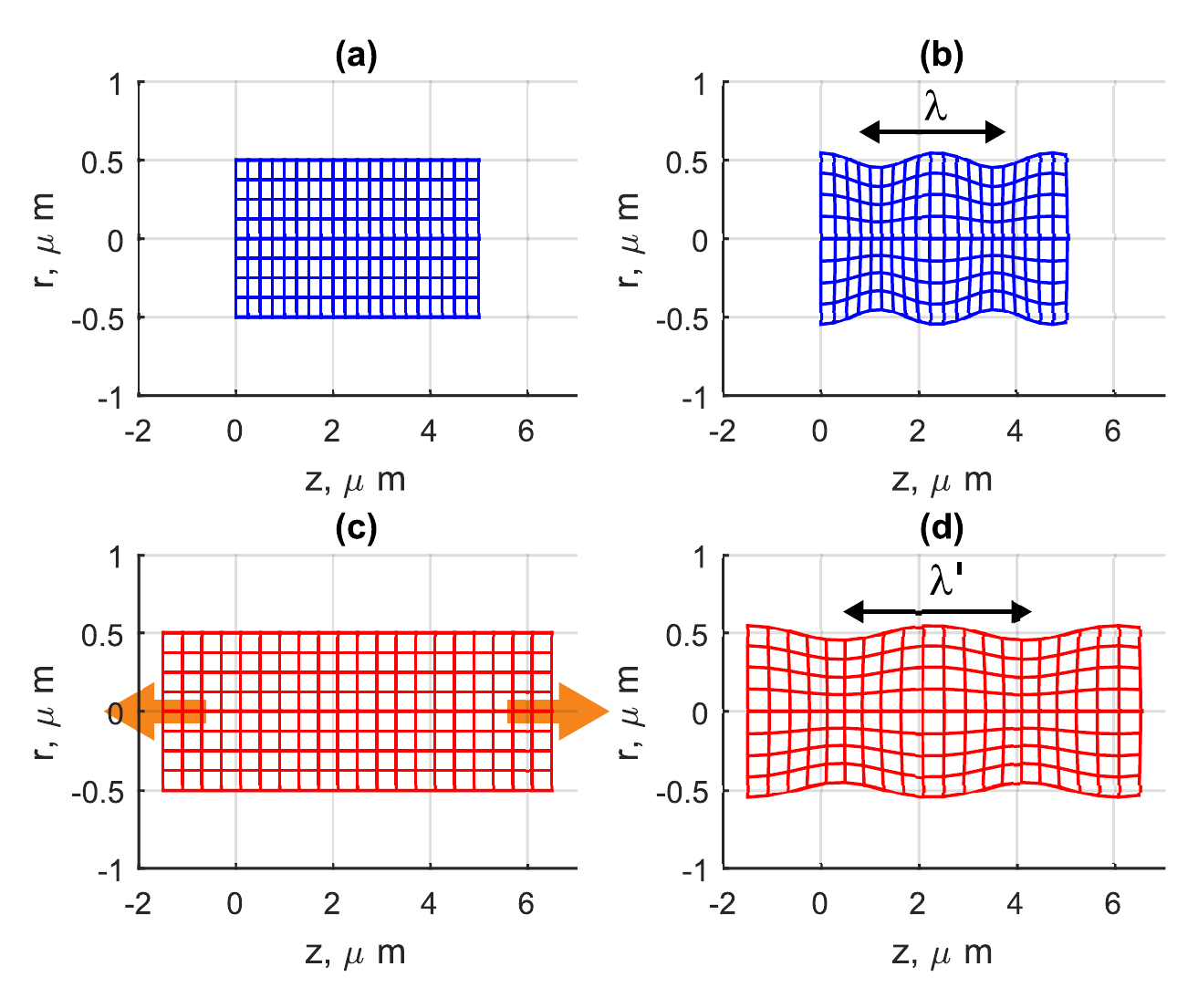}				
\caption{Schematic representation of the static and dynamic problems. (a) shows a side view of the nanofiber standing still. (b) is a view of the nanofiber supporting a surface acoustic wave. (c) is the nanofiber under longitudinal strain. The matter is stretched by the static displacement. (d) In the linear regime, the dynamic solution is the same as in (a). However the combination of static and dynamic displacement shows a longer acoustic wavelength due to the static displacement. The acoustic velocity increases accordingly.}\label{fig:static}
\end{figure}
In this section, we compute the static tensile $\bar{\epsilon}_{zz}$ induced on the nanofiber by the two translation stages in our setup. Here the static problem is reduced to a one-dimensional problem along the $z$-axis. Using the reduced elastic constant notation, the nonlinear Hook's law, given by Equation (\ref{eq:02}), can be rewritten as follows
\begin{equation}
	\bar{\sigma}_{zz}=C_{33}\bar{\epsilon}_{zz}+\frac{1}{2}C_{333}\bar{\epsilon}_{zz}^2
	=\frac{\bar{F}}{\pi r^2(z)},\label{eq:static1}
\end{equation}	
where $\bar{F}$ is the force in Newton and $r(z)$ is the taper radius at position $z$. To discriminate the strain applied to the nanofiber from the one applied to the taper transitions, we consider that the force $\bar{F}$ is uniformly distributed along the taper. The stress then varies inversely to the taper surface. So the taper profile will determine which part of the taper will be mainly strained. Actually the nanofiber is the part that experiences most strain, as previously demonstrated by Holleis \textit{et al.} \cite{Holleis_2014}. If we writes the stress in the taper waist as $\bar{T}_w=\frac{\bar{F}}{\pi r_w^2}$, we thus find
\begin{equation}
	\bar{T}_{zz}=\bar{T}_w\frac{r_w^2}{r^2(z)}\label{eq:A02}
\end{equation}
Solving Eq. (\ref{eq:static1}) yields
\begin{equation}
		\bar{\epsilon}_{zz}(z)=\frac{C_{33}}{C_{333}}\left\{-1\pm \sqrt{1+2\frac{C_{333}}{C_{33}^2}\frac{F}{\pi r^2(z)}}\right\}\label{eq:A03}
\end{equation}
As the total strain is the sum of all infinitesimal strains along $z$, for a taper of length $L$, the total elongation $\Delta L$ must satisfy
\begin{equation}
	\frac{\Delta L}{L}= \int_{-L/2}^{L/2} \bar{\epsilon}_{zz}(z) \dd z \label{eq:A04}
\end{equation}
Combining equations (\ref{eq:A03}) and (\ref{eq:A04}) we may numerically solve and retrieve the force $F$. The knowledge of the taper profile $r(z)$ enables then to compute the strain $\bar{\epsilon}_{zz}(z)$ for each section of the taper, including the nanofiber.\\

\section{Combination of both static and dynamic strains}\label{A2}
Since the nanofibers can experience strain up to few percents, the combination of both static and dynamic displacement requires careful attention. For the static problem, the nonlinear term is negligible, and we find straightforwardly from the elastodynamics equation
 \begin{eqnarray*}
	\bar{u}_{zz}(z)&=&-\Delta{l}+2\frac{\Delta{l}}{l}z,\\
\end{eqnarray*}
where $2\Delta{l}$ is the total stretching length and $l$ is the initial nanofiber length and $\bar{u}_{zz}$ the longitudinal displacement. 
Now, we must emphasize that the overall displacement field is the superposition of both the static and dynamic solution, as $u_{zz}=\bar{u}_{zz}+\tilde{u}_{zz}$. 
In Fig. (\ref{fig:static})(a), we schematically depict a side view of the nanofiber without strain. Calculating the surface acoustic wave propagating in this nanofiber, we find it is supported by a dynamic displacement field $\tilde{u}_{zz}$, as shown in Fig. (\ref{fig:static})(b). This surface wave features a wavelength $\lambda$ shown by a double black arrow. Now, as we apply a tensile strain to the nanofiber, it is stretched by a longitudinal displacement $\bar{u}_{zz}$, as shown in Fig. (\ref{fig:static})(c). Neglecting the nonlinearity we can compute the total displacement by adding both the static $\bar{u}_{zz}$ and dynamic $\tilde{u}_{zz}$. The result is shown in Fig. (\ref{fig:static})(d) that reveals that the effective acoustic wavelength $\lambda'$ is now larger to $\lambda$ owing to the strain. The vibration is still transferred from one ``column'' to the next at the same velocity, however since the all the ``columns'' are spread apart, the effective wave velocity increases. This effect is usually negligible for low strain but starts to be remarkable in our case with strain up to 6\%.

\end{document}